\pdfoutput=1
\documentclass{article}
\usepackage{filecontents}

\usepackage{geometry}
\usepackage{changepage}
\usepackage{xr-hyper}
\usepackage{hyperref}
\usepackage[utf8]{inputenc}
\usepackage{amsmath}
\usepackage{amssymb}
\usepackage{mathtools}

\usepackage[abs]{overpic}
\usepackage{subfig}

\makeatletter
\newcommand*{\addFileDependency}[1]{
  \typeout{(#1)}
  \@addtofilelist{#1}
  \IfFileExists{#1}{}{\typeout{No file #1.}}
}
\makeatother

\newcommand*{\myexternaldocument}[1]{%
    \externaldocument{#1}%
    \addFileDependency{#1.tex}%
    \addFileDependency{#1.aux}%
}

\myexternaldocument{Supplementary}

\def \ve{\varepsilon}

\begin{document}

\title{A discrete-to-continuum model for the human cornea with application to keratoconus}




\author{
J.~Köry$^{1,*}$, P.~S. Stewart$^{1}$, N.~A. Hill$^{1}$,
X.~Y. Luo$^{1}$ and A. Pandolfi$^{2}$}

\date{%
    $^1$School of Mathematics and Statistics, University of Glasgow, University Place, Glasgow G12 8QQ, UK\\
    $^{2}$Department of Civil and Environmental Engineering, Politecnico di Milano, Piazza Leonardo da Vinci 32, 20133 Milano, Italy\\
    $^*$Corresponding author: jakub.koery@glasgow.ac.uk\\[2ex]
    \today
}






\maketitle

\providecommand{\keywords}[1]{\textbf{\textit{Keywords: }} #1}

\begin{abstract}
We introduce a discrete mathematical model for the mechanical behaviour of a planar slice of human corneal tissue, in equilibrium under the action of physiological intraocular pressure (IOP). The model considers a regular (two-dimensional) network of structural elements mimicking a discrete number of parallel collagen lamellae connected by proteoglycan-based chemical bonds (crosslinks). Since the thickness of each collagen lamella is small compared to the overall corneal thickness, we upscale the discrete force balance into a continuum system of partial differential equations and deduce the corresponding macroscopic stress tensor and strain energy function for the micro-structured corneal tissue. We demonstrate that,  for physiological values of the IOP, the predictions of the discrete model converge to those of the continuum model. We use the continuum model to simulate the progression of the degenerative disease known as keratoconus, characterized by a localized bulging of the corneal shell. We assign a spatial distribution of damage (i.~e., reduction of the stiffness) to the mechanical properties of the structural elements and predict the resulting macroscopic shape of the cornea, showing that a large reduction in the element stiffness results in substantial corneal thinning and a significant increase in the curvature of both the anterior and posterior surfaces. 
\end{abstract}




\section{Introduction}

The cornea is the external lens of the eye, with specific mechanical and optical functions. The cornea confines and protects the anterior chamber, and it refracts the light rays supplying about two thirds of the total refractive power of the eye. Structurally, the cornea is a layered shell, where each layer is approximately uniformly curved, being pressurized on the posterior surface by the intraocular pressure (IOP) due to the presence of ocular fluids. The cornea comprises five main layers; from the anterior surface, the layers are the epithelium, the Bowman membrane, the stroma, the Descemet membrane, and the endothelium. The stroma is the thickest layer, playing the main structural role, and it is composed of a network of approximately equidistant-equidiameter collagen fibrils immersed in a matrix of proteoglycans, responsible for the formation of chemical bonds (crosslinks). The stroma is the core of our modelling study.

Layers of collagen fibrils, organized into ribbon-like lamellae, are interwoven in a complicated pattern to give the cornea structural integrity \cite{Meek2004,Meek2009}. 
Average mechanical properties of human corneal tissue can be estimated \emph{ex vivo} using simple experimental tests \cite{Elsheikh2007, Wang2021}, but more recent work has highlighted how material properties vary through the tissue thickness  \cite{Petsche2012,Winkler2013,White2017Elastic}. Aiming to understand how the organization of the constituents of the cornea affects the overall mechanical response, theoretical and numerical approaches have been used to introduce this information in constitutive models \cite{Elsheikh2007Numerical,Pandolfi2008, Studer2010, Petsche2013,Whitford2015Biomechanical}. In general, the human cornea is modelled as nearly-incompressible hyperelastic material with highly nonlinear behaviour. The strain energy function is decomposed into an isotropic contribution, modelling the proteoglycan ground matrix (\emph{e.g.} neo-Hookean or Mooney-Rivlin materials), and an anisotropic contribution, incorporating nonlinear strain-stiffening effects, describing preferential orientations and spatial dispersion of the collagen fibrils \cite{Holzapfel2000New,Ogden1972Large,Markert2005General}. However, continuum models are rather limited in their ability to incorporate local changes in the material properties and are thus unable to fully capture the localized degeneration of the various stromal components \cite{Pandolfi2006}. Moreover, as all non-collagenous components of the cornea are accounted for as a single (isotropic) contribution to the strain energy, the mechanical role of the crosslinked network of proteoglycans is not explicitly modelled. 


Keratoconus is a degenerative disease of the eye characterized by corneal thinning and uneven protrusion of the corneal tissue, which can lead to a loss of vision. Despite the availability of several clinical treatments (\emph{e.g.} corneal transplant \cite{Price2005Descemet}, lamellar keratoplasty \cite{Melles1999New} and crosslinking with riboflavin \cite{Raiskup2008Collagen,Baiocchi2009Corneal}), the aetiology of keratoconus is not yet fully understood. Multiple irreversible changes in the organization of the collagen architecture and in the chemical composition have been shown to accompany the progression of the disease \cite{Huang1996Histochemical,Rabinowitz1998Keratoconus,Ambekar2011Effect,Pandolfi2023}. Firstly, keratoconus is associated with a reduction in the number of collagen lamellae through the corneal thickness \cite{Patey1984,Pouliquen1987}. Secondly, in keratoconus the collagen fibril structure has been observed to become disordered, contrary to the high level of fibril organization typical of a healthy cornea \cite{Fullwood1992}. Thirdly, these changes in the collagen composition within the cornea are also mirrored by changes in the interconnecting proteoglycan bonds. For example, keratoconus is associated with a reduction in the density of keratan sulfate proteoglycans compared to a healthy cornea \cite{Wollensak1990,Sawaguchi1991} and a corresponding increase in dermatan sulfate molecules which are comparatively softer \cite{Haverkamp2005}. Several distinct perturbations (including chemical, genetic, and mechanical) have been investigated (alone or in combination) as potential causes of the disorder \cite{Pandolfi2023}, but the exact causal relationships between these changes and their impact on mechanical, geometrical, and optical properties of the cornea remain unclear. 


Theoretical models of corneal mechanics have been adapted to model the progression of keratoconus, mimicking the underlying weakening of the tissue via a localized reduction of the material stiffness. While existing continuum models have successfully predicted small modifications of the initial (nearly spherical) shape of the cornea, models have so far been unable to predict large scale conical deformations because the underlying models cannot systematically capture microstructural changes in the disease \cite{Pandolfi2006,Simonini2022}. Recent discrete models have managed to overcome this barrier, but are so far limited to two collagen layers through the corneal thickness and so cannot reliably represent the stromal microstructure \cite{Pandolfi2019, Pandolfi2023}. 

In this study we propose an extension of these discrete models which includes a large number of layers of collagen lamellae, from which we can systematically derive a continuum model capable of predicting the large scale cornea deformations evident in keratoconous.
Specifically, we propose a discrete, two-dimensional, model of the human cornea which allows us to describe progressive changes in the mechanical properties, leading to a modification of the corneal shape. The model includes an explicit representation of the collagen fibrils and of the crossslink microstructure. 
For simplicity, here we restrict our attention to a meridian slice of the corneal stroma of fixed depth. The strip is fixed at both ends to the limbus, as a segment of a cylindrical annulus in plane-strain configuration, loaded with the IOP on the posterior surface. 
Aiming at a rational continuum description of the stroma, we use discrete-to-continuum upscaling to obtain the corresponding strain-energy density \cite{Barry2022,Kory2023}, which will provide the stress and the local stiffness of the tissue.
We use this framework to model the onset and progression of keratoconus, by imposing a reduction in the stiffness of the stromal components localized at the central portion of the cornea, and compute the resulting displacement field and the corresponding stress and strain distributions.


The paper is organized as follows. In Sec.~\ref{sec:discreteModel}, we propose a new discrete model for the mechanical behaviour of a slice of human corneal tissue. In Sec.~\ref{sec:upscaling} we use discrete-to-continuum analysis to derive a corresponding macroscale continuum description of this tissue slice. In Sec.~\ref{sec:Results} we compare the outcomes of discrete and continuum models, and then apply the continuum model to investigate response of the tissue to a prescribed reduction in stiffness of both the collagen lamellae and the proteoglycan matrix, as a simple model for the formation of keratoconus. The model is critically discussed in Sec.~\ref{sec:discussion}. 

\section{Discrete model}
\label{sec:discreteModel}

\begin{figure}[!htbp]
\centering
\begin{overpic}[width=1.0\textwidth,tics=10]{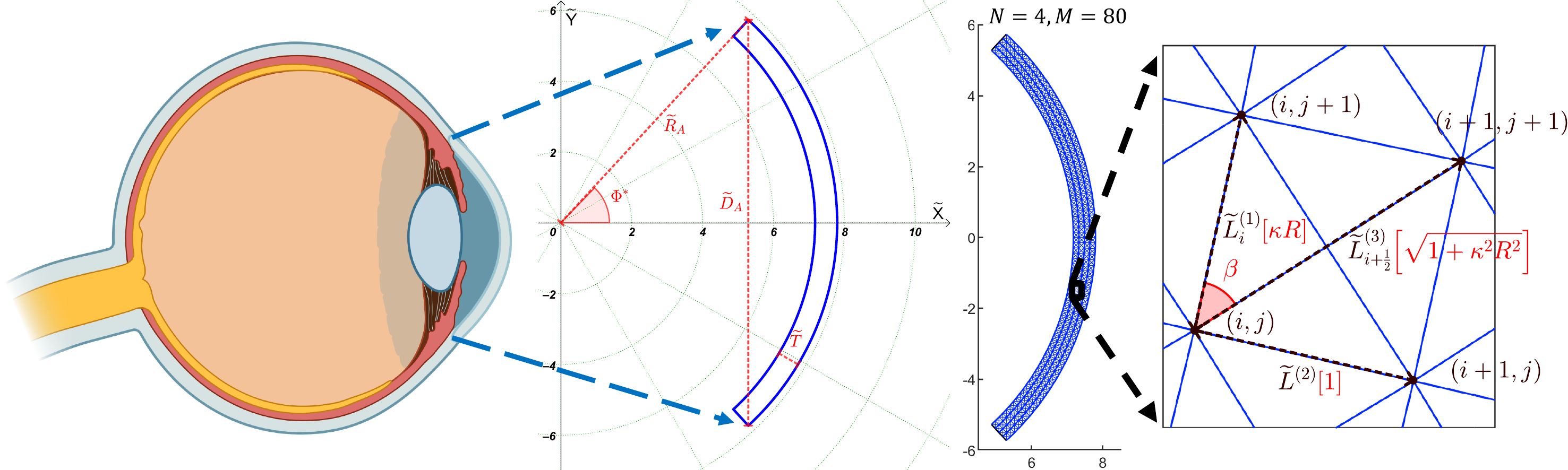}
\end{overpic}
\caption{The unloaded geometry. Zoom onto the macroscale geometry of an idealized two-dimensional corneal slice (created with BioRender.com). The unloaded-configuration cornea is shown in blue, the key parameters are indicated in red and the green dotted lines (circles) depict the curves of constant $\Phi$ ($\tilde{R}$). The unloaded configuration is then discretized ($N=4$, $M=80$, $\gamma=20$). Panel on the right presents a zoom onto the unit cell with the dimensional lengths of the elements (denoted with tilde) dependent on the radial position (index $i$). Terms in red represent corresponding quantities in the continuum limit ($N \to \infty$) of the dimensionless model. Note that while the unit cell for a finite $N$ and $M$ forms an isosceles trapezoid, in the continuum limit this becomes a rectangle.
}
\label{fig:ReferenceGeometryDiscretized}
\end{figure}

We consider a 
two-dimensional model of the human cornea, consisting of a thin meridian slice of fixed depth $\tilde{D}$ and uniform thickness $\tilde{T}$. We assume that the unloaded anterior and posterior surfaces of the cornea are concentric circular arcs, spanned by an angle $2 \Phi^*$. We introduce a planar coordinate system with origin at the centre of the circular arcs, parameterising the domain with two-dimensional Cartesian coordinates $(X,Y)$, such that the $X$-axis cuts the midpoint of the two circular arcs, see Fig.~\ref{fig:ReferenceGeometryDiscretized}. In the following, dimensional variables are denoted with the tilde, while dimensionless variable are plain. 

The geometry of the model is described by three parameters: the curvature radius of the anterior surface $\tilde{R}_A = 7.8$ mm, the uniform thickness $\tilde{T} = 0.62$ mm, and the anterior in-plane diameter $\tilde{D}_A= 11.46$ mm \cite{Pandolfi2006}. The in-plane diameter is related to the  aperture angle $2 \Phi^*$ as $\tilde{D}_A = 2 \tilde{R}_A \sin{\Phi^*}$, and it follows that $\Phi^* \approx 0.83$ rad.


We note that in plane strain configuration the depth is set to one unit of length, in this case $\tilde{D}=1$ mm. 


\begin{table}[]
    \centering
    \begin{tabular}{|ccc|cc|cccc|cc|}
         \multicolumn{3}{c}{Macroscale geometry} & \multicolumn{2}{c}{Discretization} & \multicolumn{4}{c}{Stiffnesses and BCs} & \multicolumn{2}{c}{Damage} \\
         \hline
         $\tilde{T}$ & $\tilde{R}_A$ & $\tilde{D}_A$ & $N$ & $\gamma$ & $\tilde{\alpha}_1$ & $\tilde{K}^{(2)}$ & $\tilde{K}^{(3)}$ & $\tilde{p}$ & $\xi$ & $\mathcal{D}_{max}$ \\
        \hline \hline
         (mm) & (mm) & (mm) & -- & -- & (kPa) & (N/mm) & (N/mm) & (kPa) & -- & --\\
            0.62   & 7.8  &  11.46  & 32 & 20 & 638 & 0.0072 & 0.504 & 2 & 4 & 0.99\\
         \hline
    \end{tabular}
    \caption{Independent parameters of the corneal model and their default values.}
    \label{tab:IndependentParameters}
\end{table}

\subsection{Corneal microstructure}
\label{ssec:discretization}

In the discrete approach we model the cornea as a set of $N+1$ concentric arcs, each representing an individual collagen lamella, separated by the distance $\tilde{L}^{(2)}$ 
\begin{equation}\label{eqn:FormulaL2}
\tilde{L}^{(2)} = \tilde{T}/N \, .
\end{equation}
The radius of the $i^{th}$ circular arc is 
\begin{equation}\label{eqn:FormulaRi}
\tilde{R}_i = \tilde{R}_A - \tilde{T} + i \tilde{L}^{(2)} \, , \qquad  (i=0,\dots,N) \, ,
\end{equation}
thus, the radius of the posterior surface is $\tilde{R}_0=\tilde{R}_A-\tilde{T}$ and the radius of the anterior surface is $\tilde{R}_N=\tilde{R}_A$, see Fig.~\ref{fig:ReferenceGeometryDiscretized}. Each arc is subdivided into $M$ (even) equally spaced arcs, each sweeping an angle $\phi^* = 2 \Phi^*/M$, so that
\begin{equation}
\Phi_j = (j-M/2)\phi^*, \qquad (j=0,\dots,M),
\end{equation}
and $\Phi_0 = -\Phi^\ast$ and $\Phi_M = \Phi^\ast$ identify the boundaries. The intersections between adjacent segments are defined by a set of planar points of coordinates
\begin{equation}
\tilde{X}_{i,j} = \tilde{R}_i \cos{\Phi_j}, \qquad \tilde{Y}_{i,j} =  \tilde{R}_i \sin{\Phi_j} \qquad (i=0,...,N, \quad j=0,...,M).
\end{equation}
The resulting discretization size is defined by $N$ and $M$, and the mesh aspect ratio $\gamma=M/N$ is taken as a fixed parameter. In the spirit of truss models, two adjacent nodes are connected by a straight segment, thus each collagen lamella is a piecewise linear approximation to a circular arc. We denote the lamellar segments with the superscript $^{(1)}$. The geometric/mechanical properties of the lamellar segment between the angles $\Phi_j$ and $\Phi_{j+1}$ are denoted by the subscript $i,j+1/2$. Hence, the current length of each segment is  $\tilde{l}_{i,j+1/2}^{(1)}$, while the initial length $\tilde{L}_i^{(1)}$ is independent of $j$ and proportional to the arc radius through the relation
\begin{equation}\label{eqn:FormulaL1}
\tilde{L}_i^{(1)} = 2 \tilde{R}_i \sin{\left( \frac{\Phi^*}{M} \right)}.
\end{equation}

In the human cornea, the collagen lamellae have an average thickness of about $ 2\mu$m, and the number of lamellae across the corneal thickness varies between 200 and 500 \cite{Meek2009}. 
In our model, we assume that each circular arc is representative of the behavior of several parallel lamellae. For computational reasons we limit our model to a maximum number of arcs $N=64$. The lamellae are connected to one another via a dense network of chemical bonds (crosslinks) originated by the proteoglycans.

The hydrated ground matrix of proteoglycans compressed by the IOP transmits the load to the collagen structure \cite{Alberts2017}. We model the function of the matrix by introducing a set of radial struts, denoted with the superscript $^{(2)}$. The current length of the radial struts spanning between the nodes $i$ and $i+1$ at the polar angle $\Phi_j$ is denoted as $\tilde{l}_{i+1/2,j}^{(2)}$. The corresponding initial length, $\tilde{L}^{(2)}$, Eq.~\eqref{eqn:FormulaL2}, is uniform.

Tensile stresses resulting from the action of IOP are primarily carried by collagen lamellae, whose mechanical stability is partially provided by extracellular matrix components. The sliding between lamellae is contrasted by the presence of  proteoglycans which create crosslinks between collagen fibrils \cite{Haverkamp2005}. Elastin fibres, covered with a sheath of microfibrils, may also contribute to contrast the lamellar sliding \cite{Alberts2017,White2017Structural,White2017Elastic}. 

We mimic the combined action of these elements in providing shear stiffness to the structure by introducing diagonal struts, with geometrical and mechanical properties denoted with the superscript $^{(3)}$. The node $(i,j)$ is connected diagonally to the nodes of the two adjacent layers, \emph{i.~e.}, points $i+1,j\pm1$ and $i-1,j\pm1$. The current length of the diagonal struts is denoted with $\tilde{l}_{i\pm 1/2, j \pm 1/2}^{(3)}$. The corresponding initial lengths
\begin{equation}\label{eqn:FormulaL3}
\tilde{L}_{i \pm \frac{1}{2}}^{(3)} = \sqrt{\tilde{L}_i^{(1)} \tilde{L}_{i \pm 1}^{(1)} + \left(\tilde{L}^{(2)}\right)^2},
\end{equation}
are independent of $j$ and increase from the posterior to the anterior layer, proportionally to the radius $\tilde{R}_i$ of the arc. 

We choose an aspect ratio $\gamma=M/N$ that guarantees a marked inclination of the diagonal struts, so that the corresponding axial forces contribute to contrast the sliding between the lamellae and avoid zero energy deformation modes. 

The discrete balance equations are stated under the assumption of finite kinematics, therefore the problem can lead to multiple solutions. The unknowns of the problem, thus, are either the current coordinates $\tilde{x}$, $\tilde{y}$ of the nodes or the node displacements $\tilde{u}_X$, $\tilde{u}_Y$.


\subsection{Constitutive models}
\label{ssec:constitutiveModels}

We assume that all the elements are stress-free in the unloaded configuration, disregarding the presence of pre-stresses, typical in arteries \cite{Chuong1986}. For simplicity, we assume that the force in each structural element is modelled as a linear function of its elongation, \emph{i.~e.}, the difference between the current length $\tilde{l}^{(m)}$ and the initial length $\tilde{L}^{(m)}$, $m=1,2,3$. Denoting the stretch of a structural element in the $m$ family as $\lambda^{(m)} = \tilde{l}^{(m)}/\tilde{L}^{(m)}$, the force in each element is given by 
\begin{equation}\label{eqn:LinearSpringsGeneralModel}
    \tilde{f}^{(m)}(\tilde{l}^{(m)};\tilde{L}^{(m)}) 
    = 
    \tilde{K}^{(m)} \left( \tilde{l}^{(m)} - \tilde{L}^{(m)} \right) =  \tilde{K}^{(m)} \tilde{L}^{(m)} (\lambda^{(m)} - 1) \, , 
    \quad  
    \tilde{K}^{(m)} = \frac{\tilde{E}^{(m)} \tilde{A}^{(m)}}{\tilde{L}^{(m)}} \, ,
\end{equation}
where $\tilde{K}^{(m)}$ is the strut axial stiffness (a force per unit length). The axial stiffness is expressed as the product of the Young's modulus $\tilde{E}^{(m)}$ of the material and of the cross-sectional area $\tilde{A}^{(m)}$ \cite{Gere2012mechanics}. In the Supplementary Material (Sec.~\ref{sec:EstimateStiffnessParameters}), we estimate from experimental data and from previous numerical studies the axial stiffness for each structural element. The adopted values are listed in Table \ref{tab:materialProperties}. 

Since the mesh aspect ratio $\gamma$ is constant, the length of all structural elements must scale as $O(1/N)$ in the limit as $N \to \infty$, cf. Eqs.~ \eqref{eqn:FormulaL2}-\eqref{eqn:FormulaL1}-\eqref{eqn:FormulaL3}. Also the the cross-sectional area $\tilde{A}^{(m)}$ of all structural elements must scale as $O(1/N)$ to ensure that the total volume of the discretized tissue remains constant. It follows that the elemental axial stiffness $\tilde{K}^{(m)}$ is independent of $N$. 

Finally, we note that both $\tilde{L}^{(1)}$ and $\tilde{L}^{(3)}$ increase slightly with the radius. However, this variation is small, because the thickness of cornea is small compared to the radius of curvature of the anterior surface. Therefore, in Eq.~\eqref{eqn:LinearSpringsGeneralModel} we use $\tilde{R}=\tilde{R}_A$ to estimate the stiffness for all the structural elements in the network.

\begin{table}[]
    \centering
    \begin{tabular}{lcccc}
         Component & Elastic modulus & [kPa] & Stiffness & [N/mm]\\
         \hline \hline
        Lamellar segments    & $\tilde{E}^{(1)}$ & 7656 & $\tilde{K}^{(1)}$ & 7.2 \\
        Radial crosslinks     & $\tilde{E}^{(2)}$ & 7.656    & $\tilde{K}^{(2)}$ & 0.0072 \\
        Diagonal crosslinks & $\tilde{E}^{(3)}$ & 535.92   & $\tilde{K}^{(3)}$ & 0.504\\     
        \hline
    \end{tabular}
    \caption{Material parameters used in simulations.}
    \label{tab:materialProperties}
\end{table}

\subsection{Boundary conditions}
\label{ssec:boundaryConditions}

\begin{figure}[!htbp]
\centering
\begin{overpic}[width=1.0\textwidth,tics=10]{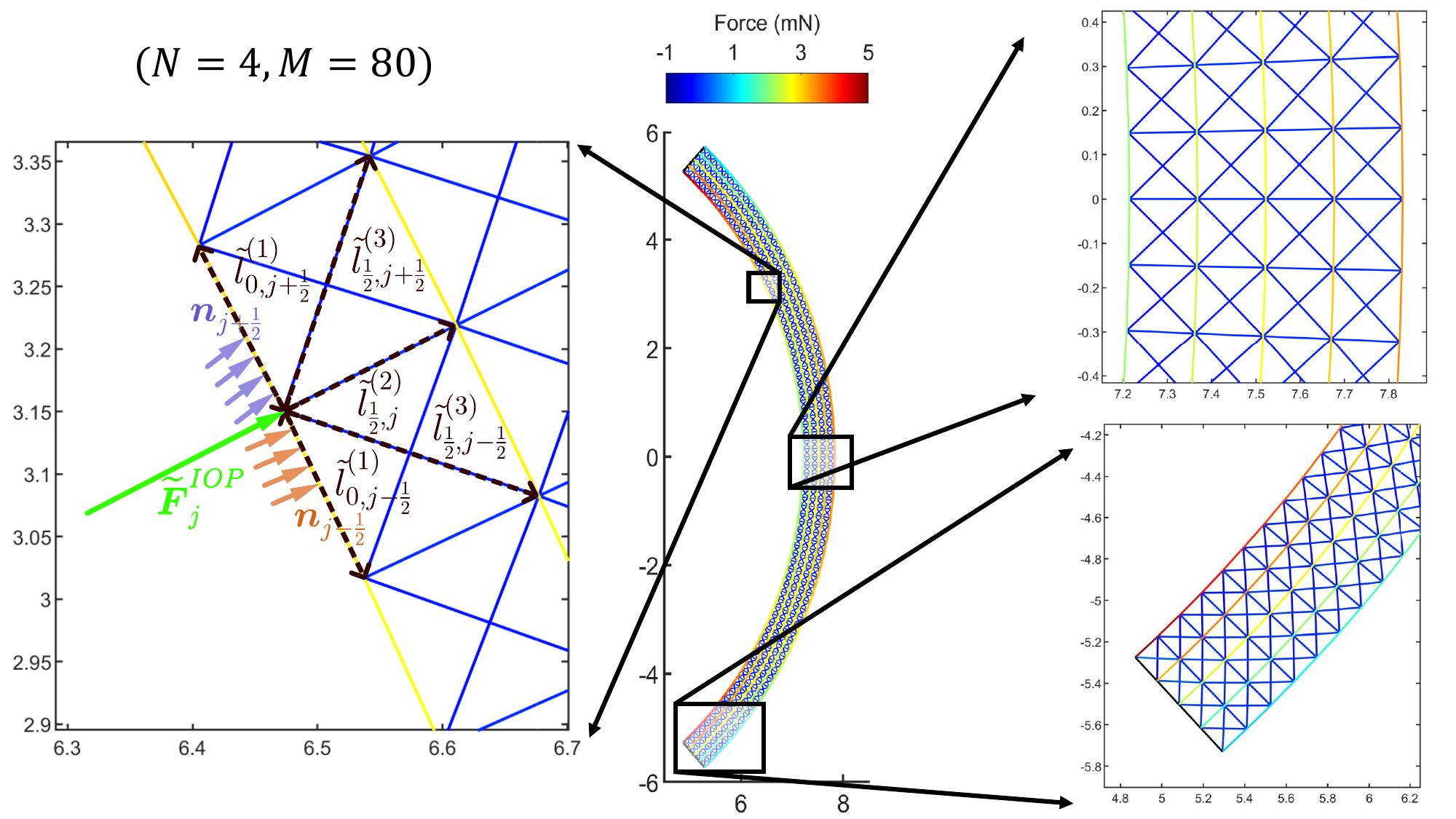}
\end{overpic}

\caption{Healthy cornea loaded with IOP. The central panel shows the loaded configuration ($\tilde{p} = 2$ kPa) of the discrete model with $N=4$ and $M=80$. Note that the colour represents the axial force in the elements with positive (negative) values indicating tension (compression). Schematic in the left panel depicts how continuum IOP is transformed into discrete forces at posterior hinges. Panels on the right present zoomed-in steady-state force distributions through the corneal thickness at the apex (top) and the limbus (bottom).}
\label{fig:DeformedGeometryDiscretized}
\end{figure}

The cornea is constrained at the two ends to the limbus, a stiff collageneous structure that contrasts the in-plane radial displacements. To mimic the limbus, we impose zero displacement for the nodes located along $\Phi=\pm\Phi^\ast$. The posterior surface is pressurized by the IOP $\tilde{p}$, while the anterior surface is traction free. The force acting on a segment of the lamellar arc between $j$ and $j+1$ on the posterior surface is computed by integrating the IOP over the surface of the segment in the form
\begin{equation}
\tilde{\boldsymbol{F}}_{j+\frac{1}{2}} =\tilde{l}_{0,j+\frac{1}{2}}^{(1)} \tilde{D} \,  \tilde{p} \, {\hat {\boldsymbol{n}}}_{j + \frac{1}{2}} ,
\end{equation}
where ${\hat {\boldsymbol{n}}}_{j+\frac{1}{2}}$ is the inward unit normal to the segment. In a trusswork only nodal loads are allowed, so the equivalent force acting on a node on the posterior surface is computed as the sum of half contributions from the two adjacent segments, see Fig.~\ref{fig:DeformedGeometryDiscretized}, as
\begin{equation}\label{eq:discreteIOP}
\tilde{\boldsymbol{F}}_j^{IOP} = \frac{\tilde{p}\tilde{D}}{2} \left( \tilde{l}_{0,j - \frac{1}{2}}^{(1)} \hat{\boldsymbol{n}}_{j - \frac{1}{2}}   + \tilde{l}_{0,j + \frac{1}{2}}^{(1)} \hat{\boldsymbol{n}}_{j + \frac{1}{2}}  \right) \, .
\end{equation}


\subsection{Discrete force balance}
\label{ssec:discreteForceBalance}

Under loading each node point moves from its initial location ${\tilde{\boldsymbol{X}}}_{i,j} = (\tilde{X}_{i,j},\tilde{Y}_{i,j})$ to a deformed location ${\tilde{\boldsymbol{x}}}_{i,j} = (\tilde{x}_{i,j},\tilde{y}_{i,j})$ generating forces in the connected rods. The force balance on each internal node of the domain includes the contribution of eight struts, in the form 
\begin{equation}\label{eqn:dimForceBalanceCircularNew}
\begin{aligned}
\boldsymbol{0} = & 
\tilde{g}^{(1)} \left( \frac{\tilde{l}_{i,j-\frac12}^{(1)}}{\tilde{L}_i^{(1)}} \right) 
\left(\tilde{\bf{x}}_{i,j-1} - \tilde{\bf{x}}_{i,j}\right) 
+ 
\tilde{g}^{(1)} \left( \frac{\tilde{l}_{i,j+\frac12}^{(1)}}{\tilde{L}_i^{(1)}} \right) 
(\tilde{\bf{x}}_{i,j+1} - \tilde{\bf{x}}_{i,j}) 
+ \\
&  
\tilde{g}^{(2)} \left( \frac{\tilde{l}_{i-\frac12,j}^{(2)}}{\tilde{L}^{(2)}} \right) 
(\tilde{\bf{x}}_{i-1,j} - \tilde{\bf{x}}_{i,j}) 
+  
\tilde{g}^{(2)} \left( \frac{\tilde{l}_{i+\frac12,j}^{(2)}}{\tilde{L}^{(2)}} \right)
(\tilde{\bf{x}}_{i+1,j} - \tilde{\bf{x}}_{i,j}) 
+ \\
&  
\tilde{g}^{(3)} \left( \frac{\tilde{l}_{i-\frac12,j-\frac12}^{(3)}}{\tilde{L}_{i-\frac{1}{2}}^{(3)}} \right) (\tilde{\bf{x}}_{i-1,j-1} - \tilde{\bf{x}}_{i,j} ) 
+  
\tilde{g}^{(3)} \left( \frac{\tilde{l}_{i-\frac12,j+\frac12}^{(3)}}{\tilde{L}_{i-\frac{1}{2}}^{(3)}} \right) (\tilde{\bf{x}}_{i-1,j+1} - \tilde{\bf{x}}_{i,j})
+ \\
&   
\tilde{g}^{(3)} \left( \frac{\tilde{l}_{i+\frac12,j-\frac12}^{(3)}}{\tilde{L}_{i + \frac{1}{2}}^{(3)}} \right)(\tilde{\bf{x}}_{i+1,j-1} - \tilde{\bf{x}}_{i,j}) 
+  
\tilde{g}^{(3)} \left( \frac{\tilde{l}_{i+\frac12,j+\frac12}^{(3)}}{\tilde{L}_{i + \frac{1}{2}}^{(3)}} \right) (\tilde{\bf{x}}_{i+1,j+1} - \tilde{\bf{x}}_{i,j}) \, .
\end{aligned}
\end{equation}
To facilitate the upscaling procedure described in Sec.~\ref{sec:upscaling}, we have introduced the force per unit length $\tilde{g}^{(m)}$, $m=1,2,3$, as a function of the stretch, in the form
\begin{equation}\label{eqn:defineStiffnessNew}
\tilde{g}^{(m)}  = 
\frac{1}{\tilde{l}} \tilde{f}^{(m)} \left( \frac{\tilde{l}}{\tilde{L}} \right) \, = \tilde{K}^{(m)} \left(1- \lambda^{-1} \right) \, . 
\end{equation}

Clearly, each node at the posterior and anterior surfaces of the cornea receives the contribution from five struts. The nodes on the posterior surface are loaded with the IOP, Eq.~\eqref{eq:discreteIOP}. The nodes on the limbus are fixed and they are excluded from the balance equations.

\subsection{Nondimensionalization}

We formulate the problem in non-dimensional form as follows. We proceed by dividing all the lengths by the initial corneal thickness $\tilde{T}$ and all forces by the ``reference'' force in the lamellar segments $\tilde{K}^{(1)} {\tilde T}$. The resulting model is governed by seven dimensionless quantities
\begin{equation}
\gamma = \frac{M}{N}, \quad \Phi^\ast, \quad R_A= \frac{\tilde{R}_A}{\tilde{T}}, \quad D =\frac{\tilde{D}}{\tilde{T}}, \quad K^{(2)} = \frac{{\tilde K}^{(2)}}{{\tilde K}^{(1)}}, \quad K^{(3)} =\frac{{\tilde K}^{(3)}}{{\tilde K}^{(1)}}, \quad p =\frac{{\tilde p} {\tilde T}}{{\tilde K}^{(1)}},
\end{equation}
where $\gamma$ defines the mesh aspect ratio, $\phi^\ast$ the angle swept out from limbus to limbus, $R_A$ the radius of the anterior surface, $D$ the ratio between depth and thickness, $K^{(2)}$ and $K^{(3)}$ the stiffness of the radial and of the diagonal elements, and $p$ the IOP, respectively. 

For convenience, we introduce the dimensionless ratios between the initial lengths of the structural elements as
\begin{equation}
\label{eq:qs}
q_i  \equiv \frac{\tilde{L}^{(2)}}{\tilde{L}_i^{(1)}} = \frac{\tilde{T}}{2 N \tilde{R}_i \sin{\left( \Phi^* / M \right)}} , \qquad w_{i \pm \frac{1}{2}} \equiv \frac{\tilde{L}^{(2)}}{\tilde{L}_{i \pm \frac{1}{2}}^{(3)}} =  \frac{\sqrt{q_i q_{i \pm 1}}}{\sqrt{1+q_i q_{i \pm 1}}}. 
\end{equation}
From Eq.~\eqref{eqn:defineStiffnessNew} we obtain the dimensionless force per unit length in each strut as
\begin{equation}\label{eqn:LinearSpringsFormOfStiffnessesDimless}
g^{(m)}(\lambda) = K^{(m)} \left(1- \lambda^{-1} \right), \qquad m=1,2,3.
\end{equation}

The dimensionless balance equations of the system are rendered as follows. For the interior points of the domain ($i=1,...,N-1$ and $j=1,...,M-1$), the  balance equation Eq.~\eqref{eqn:dimForceBalanceCircularNew} takes the form
\begin{subequations}\label{eqn:dimLessFullProblem}
\begin{equation}\label{eqn:dimLessForceBalanceCircularNew}
\begin{aligned}
0 =  & g^{(1)} \left(q_i N l_{i,j-\frac12} \right) ({\bf {x}}_{i,j-1} - {\bf {x}}_{i,j} )  + 
g^{(1)} \left(q_i N l_{i,j+\frac12} \right) ({\bf {x}}_{i,j+1} - {\bf {x}}_{i,j} ) \\
& + g^{(2)} \left(N l_{i-\frac12,j} \right) ({\bf {x}}_{i-1,j} - {\bf {x}}_{i,j})  +
g^{(2)} \left(N l_{i+\frac12,j} \right) ({\bf {x}}_{i+1,j} - {\bf {x}}_{i,j}) \\
& + g^{(3)} \left( w_{i-\frac12} N  l_{i-\frac12,j-\frac12} \right) ({\bf {x}}_{i-1,j-1} - {\bf {x}}_{i,j} )  \\
& +  g^{(3)} \left(w_{i-\frac12} N l_{i-\frac12,j+\frac12} \right) ({\bf {x}}_{i-1,j+1} - {\bf {x}}_{i,j} ) \\
& + g^{(3)} \left( w_{i+\frac12} N l_{i+\frac12,j-\frac12} \right) ({\bf {x}}_{i+1,j-1} - {\bf {x}}_{i,j} ) \\
& +  g^{(3)} \left(w_{i+\frac12} N l_{i+\frac12,j+\frac12} \right) ({\bf {x}}_{i+1,j+1} - {\bf {x}}_{i,j} ) \, .
\end{aligned}
\end{equation}
The boundary conditions at the limbus ($i=0,...,N$ and $j=0 \text{ or }M$) become
\begin{equation}\label{eqn:dimLessLimbusFixed}
{\bf{x}}_{i,0} = {\bf{X}}_{i,0} \qquad {\bf{x}}_{i,M} = {\bf{X}}_{i,M} \, .
\end{equation}
The nodes on the posterior surface ($j=1,...,M-1$ and $i=0$) satisfy the equation
\begin{equation}\label{eqn:dimLessPosterior}
\begin{aligned}
0 =  & g^{(1)} \left(q_0 N l_{0,j-\frac12} \right) ({\bf {x}}_{0,j-1} - {\bf {x}}_{0,j} )  + 
g^{(1)} \left(q_0 N l_{i,j+\frac12} \right) ({\bf {x}}_{0,j+1} - {\bf {x}}_{0,j} ) \\
& + g^{(2)} \left(N l_{\frac12,j} \right) ({\bf {x}}_{1,j} - {\bf {x}}_{0,j}) + g^{(3)} \left( w_{\frac12} N l_{\frac12,j-\frac12} \right) ({\bf {x}}_{1,j-1} - {\bf {x}}_{0,j} ) \\
& + g^{(3)} \left(w_{\frac12} N l_{\frac12,j+\frac12} \right) ({\bf {x}}_{1,j+1} - {\bf {x}}_{0,j} ) + \frac{p D}{2} \left( l_{0,j - \frac{1}{2}}^{(1)} \boldsymbol{n}_{j - \frac{1}{2}}   + l_{0,j + \frac{1}{2}}^{(1)} \boldsymbol{n}_{j + \frac{1}{2}}  \right) \, .
\end{aligned}
\end{equation}
The nodes on the anterior surface ($j=1,...,M-1$ and $i=N$) satisfy the equation
\begin{equation}\label{eqn:dimLessAnterior}
\begin{aligned}
0 =  & g^{(1)} \left(q_N N l_{N,j-\frac12} \right) ({\bf {x}}_{N,j-1} - {\bf {x}}_{N,j} )  + 
g^{(1)} \left(q_N N l_{N,j+\frac12} \right) ({\bf {x}}_{N,j+1} - {\bf {x}}_{N,j} ) \\
& + g^{(2)} \left(N l_{N-\frac12,j} \right) ({\bf {x}}_{N-1,j} - {\bf {x}}_{N,j})  \\
& + g^{(3)} \left( w_{N-\frac12} N  l_{N-\frac12,j-\frac12} \right) ({\bf {x}}_{N-1,j-1} - {\bf {x}}_{N,j} )  \\
& +  g^{(3)} \left(w_{N-\frac12} N l_{N-\frac12,j+\frac12} \right) ({\bf {x}}_{N-1,j+1} - {\bf {x}}_{N,j} ) \, .
\end{aligned}
\end{equation}
\end{subequations}

\section{Upscaling to a continuum model}
\label{sec:upscaling}

Clearly, the discrete model becomes computationally intractable as $N$ becomes large, as we will show in Sec.~\ref{sec:Results}. We investigate the possibility to reach the continuum limit as the number of lamellae becomes large, \emph{i.~e.}, we look at the asymptotic limit $N \to \infty$ while holding the mesh aspect ratio $\gamma$ fixed. By introducing $\ve = 1/N$ as a small parameter and expanding Eq.~\eqref{eq:qs} in the limit $N \rightarrow \infty$, we obtain that Eq.~\eqref{eq:qs} becomes
\begin{equation}
q_i = \frac{\gamma \tilde{T}}{2 \Phi^* \tilde{R}_i} + O(\varepsilon^2), \qquad w_{i\pm \frac12} =  \frac{\gamma \tilde{T}}{\sqrt{(2 \Phi^*)^2 \tilde{R}_i \tilde{R}_{i \pm 1} + (\gamma \tilde{T})^2 }} + O(\varepsilon^2)  . 
\end{equation}
Technical details of the upscaling procedure are provided in Sec. \ref{app:upscaling} of the Supplementary Material. Here we summarize the key concepts. The discrete node coordinates ($x_{i,j},y_{i,j}$) are mapped to the position functions $x(X,Y)$ and $y(X,Y)$ such that
\begin{equation}
x_{i,j} = x(X=X_{i,j},Y=Y_{i,j}), \qquad y_{i,j} = y(X=X_{i,j},Y=Y_{i,j}). 
\end{equation}
Similarly, the discrete lengths $q_i$ and $w_{i \pm \frac12}$ defined in Eq.~\eqref{eq:qs} can be mapped to continuum functions of the Cartesian coordinates $(X,Y)$. However, in consideration of the circular geometry, it is more convenient to express the lengths in terms of the radial coordinate $R=\sqrt{X^2+Y^2}$, such that
\begin{equation}\label{eqn:DefsOf_q_kappa_w}
q(R) = \frac{1}{\kappa R} \, , \qquad w(R) = \frac{q}{\sqrt{1+q^2}} = \frac{1}{\sqrt{1+ (\kappa R)^2}} \, ,
\qquad \kappa = \frac{2 \Phi^*}{\gamma} \,.
\end{equation}
In the continuum limit, we define the angle $\beta(R)$ between the diagonal crosslinks and the lamellae as
\begin{equation}
\beta(R) = \tan^{-1}({q(R)}) \, ,
\end{equation}
see Fig.~\ref{fig:ReferenceGeometryDiscretized}.

Next, we use Taylor expansion to relate all the quantities contributing to the discrete force balance at node ($i,j$), see Eq.~\eqref{eqn:dimLessForceBalanceCircularNew}. In the expansion, the differences between adjacent nodes are mapped into spatial derivatives and we obtain, at $O(\ve^2)$, the continuum form of the balance equation as
\begin{equation}\label{eqn:UpscaledProblem1}
\begin{aligned}
0 = & \left( g^{(2)} \left( \sqrt{x_R^2 + y_R^2} \right) \left(x_R,y_R \right) + g^{(3)} \left( w \sqrt{ \left(  x_R + \kappa x_\Phi \right)^2 + \left( y_R + \kappa y_\Phi \right)^2} \right) \left(  x_R + \kappa x_\Phi, y_R + \kappa y_\Phi \right)   \right. \\
& \left. +  g^{(3)} \left( w \sqrt{ \left(  x_R - \kappa x_\Phi \right)^2 + \left( y_R - \kappa y_\Phi \right)^2} \right) \left(  x_R - \kappa x_\Phi, y_R - \kappa y_\Phi \right)   \right)_R \\
& + \kappa \left(  g^{(1)} \left(q \sqrt{(\kappa x_\Phi)^2 + (\kappa y_\Phi)^2} \right) \left(\kappa x_\Phi, \kappa y_\Phi \right) \right. \\
& + \left. g^{(3)} \left( w \sqrt{ \left(  x_R + \kappa x_\Phi \right)^2 + \left( y_R + \kappa y_\Phi \right)^2} \right) \left(  x_R + \kappa x_\Phi, y_R + \kappa y_\Phi \right)  \right. \\
& - \left.  g^{(3)} \left(w \sqrt{ \left(  x_R - \kappa x_\Phi \right)^2 + \left( y_R - \kappa y_\Phi \right)^2} \right) \left(  x_R - \kappa x_\Phi, y_R - \kappa y_\Phi \right)  \right)_\Phi .
\end{aligned}
\end{equation}

Finally, to facilitate the interpretation of the obtained equations in the circular geometry, we switch from Cartesian to polar coordinates, $(r,\phi)$ and $(\hat{\boldsymbol{r}}, \hat{\boldsymbol{\phi}})$ and obtain the continuum balance equation at the macroscale in the form
\begin{equation}\label{eqn:ContinuumProblemCurrentUnitVectors}
\begin{aligned}
\boldsymbol{0} = & \left\{ \kappa^2 \left( \left( r_\Phi \mathcal{G}^{(1)} \right)_{\Phi} - r \phi_\Phi^2 \mathcal{G}^{(1)} \right) +  \left( \left(r_R \mathcal{G}^{(2)} \right)_R  - r \phi_R^2 \mathcal{G}^{(2)}   \right) +  \right. \\
& \left. \left( \left( r_R + \kappa r_\Phi \right) \mathcal{G}_{+}^{(3)} + \left(r_R - \kappa r_\Phi \right) \mathcal{G}_{-}^{(3)} \right)_R + \kappa \left( \left( r_R + \kappa r_\Phi \right) \mathcal{G}_{+}^{(3)} - \left(r_R - \kappa r_\Phi \right) \mathcal{G}_{-}^{(3)}  \right)_\Phi - \right. \\
& \left. \left(\phi_R + \kappa \phi_\Phi \right)  \left( r \phi_R + \kappa r \phi_\Phi \right) \mathcal{G}_{+}^{(3)} - \left( \phi_R - \kappa \phi_{\Phi} \right) \left(r \phi_R - \kappa r \phi_\Phi \right) \mathcal{G}_{-}^{(3)} \right\} \hat{\boldsymbol{r}} + \\
& \left\{ \kappa^2 \left( \left( r \phi_\Phi \mathcal{G}^{(1)}\right)_{\Phi} + r_\Phi \phi_\Phi \mathcal{G}^{(1)} \right) +  \left( \left(r \phi_R \mathcal{G}^{(2)} \right)_R + r_R \phi_R \mathcal{G}^{(2)}   \right) + \right. \\
& \left. \left( \left( r \phi_R + \kappa r \phi_\Phi \right) \mathcal{G}_{+}^{(3)} + \left(r \phi_R - \kappa r \phi_\Phi \right) \mathcal{G}_{-}^{(3)} \right)_R  + \kappa \left( \left( r \phi_R + \kappa r \phi_\Phi \right) \mathcal{G}_{+}^{(3)} - \left(r \phi_R - \kappa r \phi_\Phi \right) \mathcal{G}_{-}^{(3)}  \right)_\Phi + \right. \\
& \left. \left( \phi_R + \kappa \phi_\Phi \right)  \left( r_R + \kappa r_\Phi \right) \mathcal{G}_{+}^{(3)} +  \left(\phi_R - \kappa \phi_{\Phi} \right) \left(r_R - \kappa r_\Phi \right) \mathcal{G}_{-}^{(3)} \right\} \hat{\boldsymbol{\phi}},
\end{aligned}
\end{equation}
where
\begin{equation}\label{eqn:DefsOfMathcalK}
\begin{aligned}
\mathcal{G}^{(1)}(R,\Phi) =  g^{(1)} &\left( q \sqrt{(\kappa r_\Phi)^2 + (\kappa r \phi_\Phi)^2}\right) \, , \qquad  \mathcal{G}^{(2)}(R,\Phi) =  g^{(2)}\left(\sqrt{r_R^2 + (r \phi_R)^2}\right) \, , \\
& \mathcal{G}_{\pm}^{(3)}(R,\Phi)  =  g^{(3)}\left(w \sqrt{(r_R \pm \kappa r_\Phi)^2+(r \phi_R \pm \kappa r \phi_\Phi)^2}\right) \, .
\end{aligned}
\end{equation}

Under the assumption of elasticity (reversibility), we derive the corresponding upscaled strain energy function of the stroma, by pulling the vectors back to the unloaded configuration, Eq.~\eqref{eqn:AlmostForceBalanceContinuum}. The local linear momentum balance in the reference configuration is expressed in terms of the second Piola--Kirchhoff stress tensor and is directly linked to the displacement field through the stretch. The symmetric second Piola-Kirchhoff stress tensor in the polar coordinates is
\begin{subequations}\label{eqn:SecondPiolaKirchhoff}
\begin{equation}
\tilde{\boldsymbol{T}}=  \left[\begin{array}{cc} 
  \tilde{T}_{RR} &  \tilde{T}_{R \Phi} \\
 \tilde{T}_{R \Phi} & \tilde{T}_{\Phi \Phi}
\end{array}\right], \qquad \text{where}
\end{equation}
\begin{equation}
\begin{aligned}
\tilde{T}_{RR} & = \frac{\tilde{K}^{(2)} \left( 1 - \left( \hat{\boldsymbol{e}}_R \cdot \left(\boldsymbol{C} \hat{\boldsymbol{e}}_R \right)  \right)^{-1/2} \right)  + \tilde{K}^{(3)} \left( 2 - \left( \hat{\boldsymbol{e}}_+ \cdot \left(\boldsymbol{C} \hat{\boldsymbol{e}}_+ \right)\right)^{-1/2} - \left( \hat{\boldsymbol{e}}_- \cdot \left(\boldsymbol{C} \hat{\boldsymbol{e}}_- \right)\right)^{-1/2} \right)}{\kappa R \tilde{D}}  \\
\tilde{T}_{R \Phi} & = \frac{\tilde{K}^{(3)} \left( \left( \hat{\boldsymbol{e}}_- \cdot \left(\boldsymbol{C} \hat{\boldsymbol{e}}_- \right)\right)^{-1/2} - \left( \hat{\boldsymbol{e}}_+ \cdot \left(\boldsymbol{C} \hat{\boldsymbol{e}}_+ \right) \right)^{-1/2} \right)}{\tilde{D}}  \\
\tilde{T}_{\Phi \Phi} & = \frac{\kappa R \left(  \tilde{K}^{(1)} \left(1- \left(\hat{\boldsymbol{e}}_\Phi \cdot \left(\boldsymbol{C} \hat{\boldsymbol{e}}_\Phi \right) \right)^{-1/2} \right) + \tilde{K}^{(3)} \left( 2 - \left( \hat{\boldsymbol{e}}_+ \cdot \left(\boldsymbol{C} \hat{\boldsymbol{e}}_+ \right)\right)^{-1/2} - \left( \hat{\boldsymbol{e}}_- \cdot \left(\boldsymbol{C} \hat{\boldsymbol{e}}_- \right)\right)^{-1/2} \right)  \right) }{\tilde{D}}  
\end{aligned}
\end{equation}
\end{subequations}
From the second Piola-Kirchhoff stress stress and its work conjugate deformation tensor it is possible to derive the the strain-energy density of the upscaled material, see Sec.~\ref{app:upscaling} of the Supplementary Material for the derivation. Specifically, the strain-energy density function has the form
\begin{equation}\label{eqn:StrainEnergyLinearSprings}
\begin{aligned}
\tilde{W} = 
& \frac{1}{2 \kappa R \tilde{D}}  \left\{ \left( \kappa R \right)^2 \tilde{K}^{(1)} \left( \sqrt{ \hat{\boldsymbol{e}}_\Phi \cdot \left(\boldsymbol{C} \hat{\boldsymbol{e}}_\Phi \right) } - 1 \right)^2 + \tilde{K}^{(2)}  \left(\sqrt{ \hat{\boldsymbol{e}}_R \cdot \left(\boldsymbol{C} \hat{\boldsymbol{e}}_R \right) } -1 \right)^2  \right. \\
& + \left. \left[ 1 + \left( \kappa R \right)^2 \right] \tilde{K}^{(3)} \left[ \left( \sqrt{ \hat{\boldsymbol{e}}_+ \cdot \left(\boldsymbol{C} \hat{\boldsymbol{e}}_+ \right) } -1 \right)^2 +           \left(\sqrt{ \hat{\boldsymbol{e}}_- \cdot \left(\boldsymbol{C} \hat{\boldsymbol{e}}_- \right) } -1 \right)^2 \right] \right\},
\end{aligned}
\end{equation}
where $\boldsymbol{C}$ is the right Cauchy-Green tensor and $\hat{\boldsymbol{e}}_{R}$, $\hat{\boldsymbol{e}}_{\Phi}$ and $\hat{\boldsymbol{e}}_{\pm}$ are the unit normal vectors pointing along the directions of the radial crosslinks, collagen lamellar segments and the diagonal crosslinks, respectively.  This result is consistent with the general form of an elasticity tensor for anisotropic linearly elastic material with two orthogonal lines of symmetry \cite{Trageser2019}, which is obtained in the continuum limit of the discrete geometry, \emph{i.~e.}, 
$$ \frac{\tilde{L}_{i+1}^{(1)}}{\tilde{L}_{i}^{(1)}} = \frac{\tilde{R}_{i+1}}{\tilde{R}_i} = 1 + \frac{\varepsilon \tilde{T}}{ \tilde{R}_i} \to 1 \qquad \text{as} \quad \varepsilon \to 0,$$
where we have used Eqs. \eqref{eqn:FormulaL1}, \eqref{eqn:FormulaRi} and \eqref{eqn:FormulaL2}, see Sec.~\ref{sec:Appendix_StrainEnergySmallDef} in the Supplementary Material. 

\section{Results}
\label{sec:Results}

In the simulations of discrete and continuum models, we assume an unloaded cornea in the reference configuration. The structure is loaded up to 2 kPa IOP ($\approx$15 mmHg, the average value in human \cite{Pandolfi2006}). The baseline parameters are listed in Table \ref{tab:IndependentParameters}. 

We solve the discrete problem on a standard desktop computer for $N$ ranging from $2$ to $64$, by using the Newton's method implemented in {\sc MATLAB} ({\tt fsolve} tool) and adopting default values of error tolerances (StepTolerance = FunctionTolerance = $10^{-6}$). Computational times range from 30 s for $N=4$ to 1 week for $N=64$. Simulations for $N=64$ become too expensive.

We solve the continuum problem by a finite element package (FEniCS, \cite{Logg2012}) with standard bilinear Lagrange elements, and a mesh discretized with $64$ elements in the radial direction and $20 \times 64 = 1,280$ elements in the meridian direction. Varying the number of elements in the radial direction from $32$ to $128$ we found that the percentage change in the apex displacement (compared to the baseline case, \emph{i.~e.} $64$ elements) was less than $0.01$ \%. 
The finite element code requires the definition of the strain energy \eqref{eqn:StrainEnergyLinearSprings} stored in the domain and of the extra energetic contribution for the intraocular pressure (as described in \cite{Logg2012}). The baseline continuum simulation with $64$ elements in the radial direction runs for less than 90 s and shows a significant improvement on the computational time of the discrete model.

We first consider the deformation of a healthy cornea (\emph{i.~e.} baseline parameter values) as it is loaded with IOP, Sec.~\ref{ssec:healthyCornea}. We then simulate damage to the corneal structure and use the continuum model to predict the corresponding macroscale deformation as might be expected during keratoconus, Sec.~\ref{ssec:keratoconus}.  

\subsection{A healthy cornea}
\label{ssec:healthyCornea}

\begin{figure}[!htbp]
    \centering
    \includegraphics[width=0.9\textwidth]{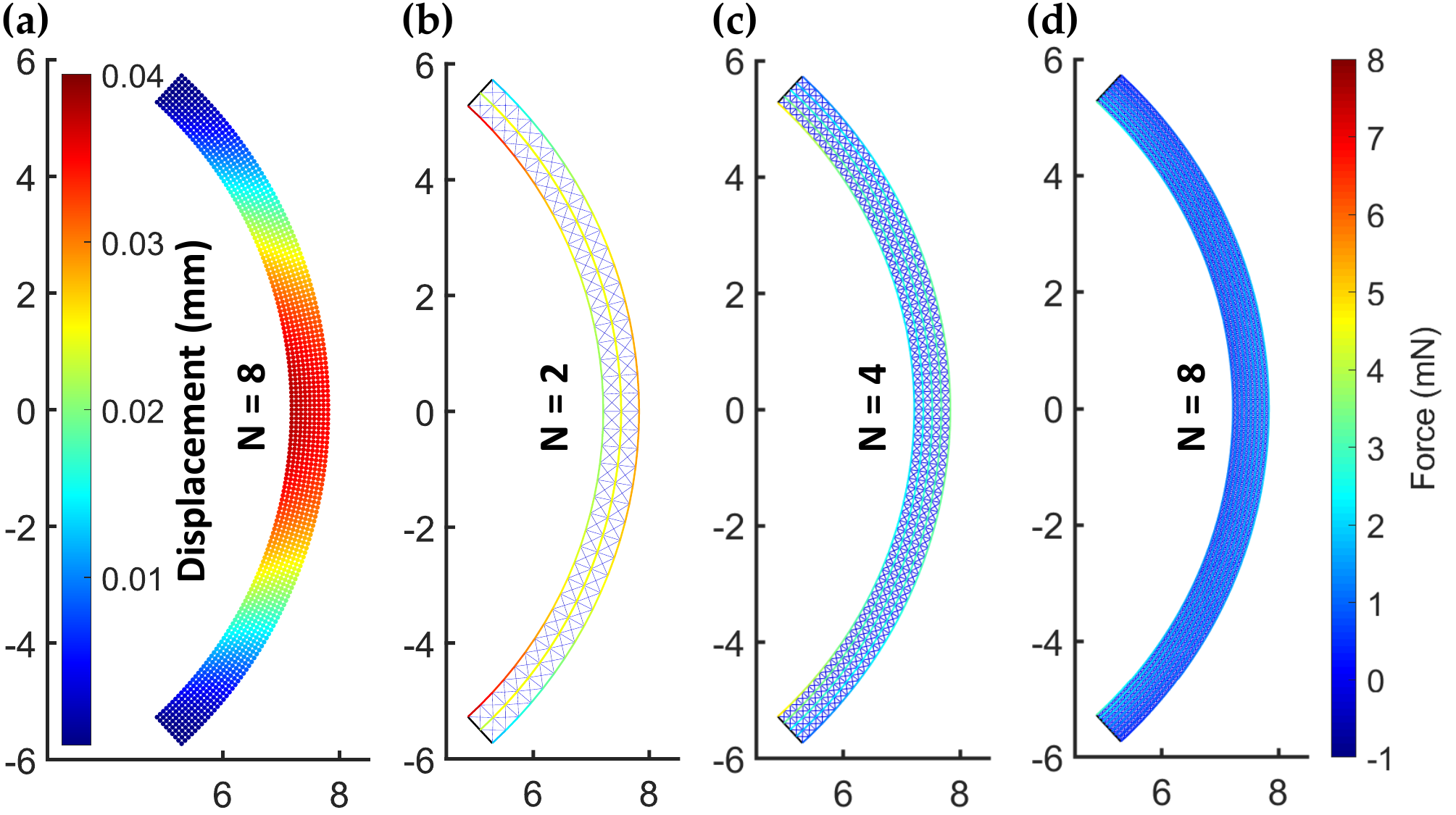}
    \includegraphics[width=1.0\textwidth]{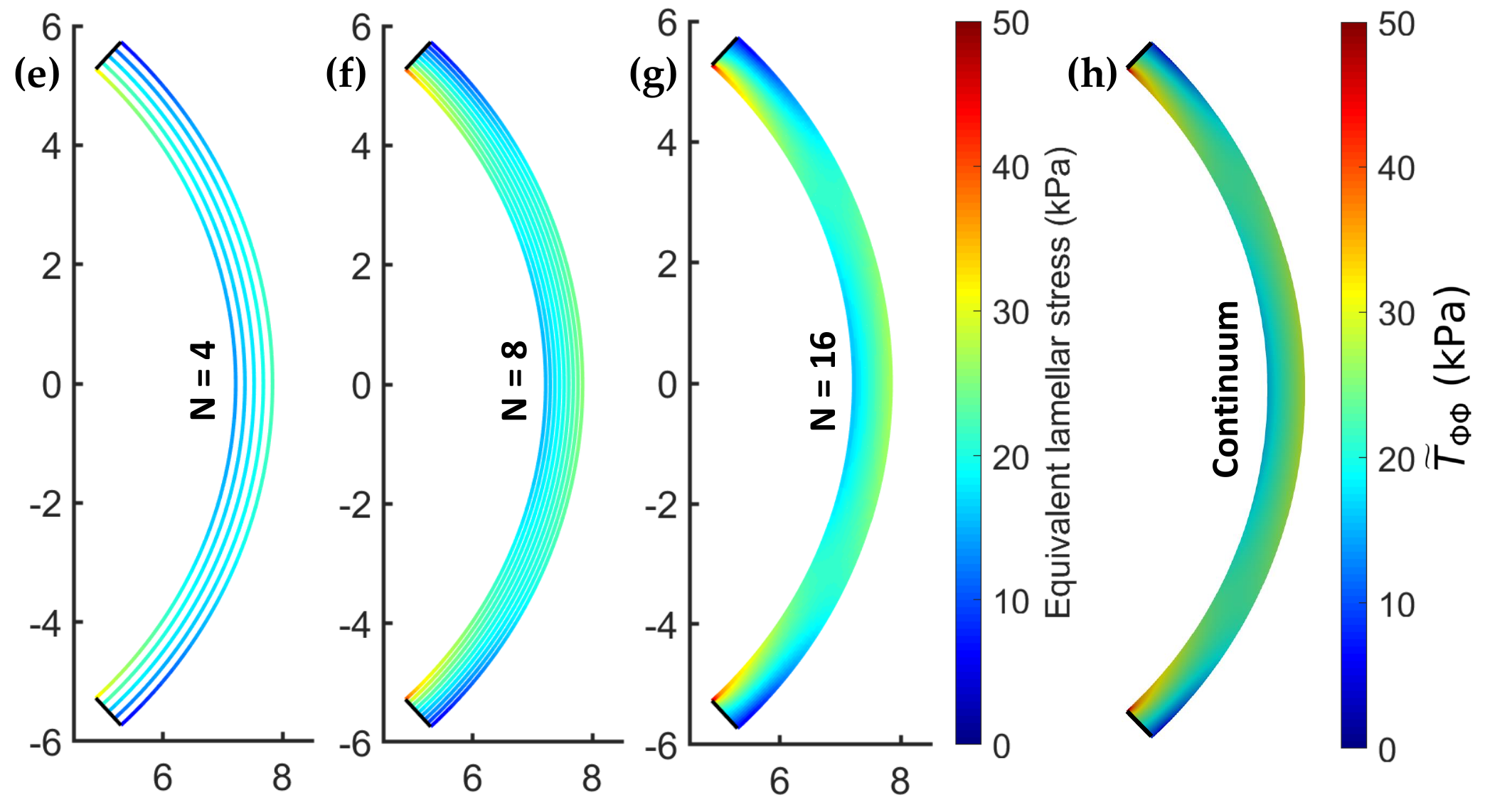}
    \caption{For a healthy cornea under physiological IOP ($2$ kPa), node displacements are small (a) and the axial forces in individual elements decrease as $N \to \infty$ (b-d). The discrete axial forces in the lamellae can be divided by an appropriate cross-sectional area to get equivalent lamellar stresses. These stresses for large enough $N$ (e-g) agree well with the circumferential stress $\widetilde{T}_{\Phi \Phi}$ in the continuum limit (h), because the lamellar segments are the stiffest elements. Note that while $\widetilde{T}_{\Phi \Phi}$ in the continuum model exceeds the displayed range in small regions near domain corners, we for ease of comparison restrict the colorbar to the same range as for the discrete model.}
    \label{fig:DiscreteToContinuum_ForcesToStress}
\end{figure}

In order to elucidate how the predictions of the discrete model depend on the number of layers of collagen lamellae, in Fig.~\ref{fig:DiscreteToContinuum_ForcesToStress} we show the results of the simulations for the discrete system  Eq.~\eqref{eqn:dimLessFullProblem} for increasing $N$ keeping the mesh aspect ratio and elastic stiffness fixed to the baseline values. 
The displacements are small, the maximum displacement at the apex being $\approx 0.04$ mm (Fig.~\ref{fig:DiscreteToContinuum_ForcesToStress}a). Furthermore, the deflections of the posterior and anterior surfaces are very small, and so their curvature remains almost uniform. The corresponding force in the structural elements is shown for a baseline simulation with  $N=2$ (Fig.~\ref{fig:DiscreteToContinuum_ForcesToStress}b), $N=4$ (Fig.~\ref{fig:DiscreteToContinuum_ForcesToStress}c) and $N=8$ (Fig.~\ref{fig:DiscreteToContinuum_ForcesToStress}d), which corresponds to the displacement profile shown in Fig.~\ref{fig:DiscreteToContinuum_ForcesToStress}a. As the number of layers is increased, the equivalent IOP force is distributed over an increasing number of structural elements decreasing their internal force  (Figs.~\ref{fig:DiscreteToContinuum_ForcesToStress}b-d). 

Within the discrete model we estimate the azimuthal component of the normal stress vector along each lamellar segment as
\begin{equation}
\tilde{T}_{\Phi, \Phi; i, j \pm \frac12} = \frac{\tilde{K}^{(1)} \left( \tilde{l}_{i, j \pm \frac12}^{(1)} - \tilde{L}_i^{(1)} \right)}{\tilde{L}^{(2)} \tilde{D}},
\end{equation}
where the force exerted on the segments is divided by their cross-sectional area in the reference configuration. This component of lamellar stress is plotted for each structural element in discrete simulations with $N=4$ layers (Fig.~\ref{fig:DiscreteToContinuum_ForcesToStress}e), $N=8$ layers (Fig.~\ref{fig:DiscreteToContinuum_ForcesToStress}f) and $N=16$ (Fig.~\ref{fig:DiscreteToContinuum_ForcesToStress}g). A consistent limit emerges as $N$ increases, where close to the centre of the cornea the lamellar stress is close to uniform and (approximately) independent of the radial coordinate (Fig.~\ref{fig:DiscreteToContinuum_ForcesToStress}f,g). 
However, the lamellar stress is very large (small) on the posterior (anterior) surface of the cornea adjacent to the limbus, leading to large gradients in stress in the radial direction, revealing the presence of a bending moment due to the rigid boundary constraints. 
Finally, we note that the lamellar stress distribution predicted by the discrete model (Fig.~\ref{fig:DiscreteToContinuum_ForcesToStress}g) approaches the corresponding prediction of the $\tilde{T}_{\Phi \Phi}$ component of the stress tensor $\tilde{\boldsymbol{T}}$ as defined in Eq.~\eqref{eqn:SecondPiolaKirchhoff} (Fig.~\ref{fig:DiscreteToContinuum_ForcesToStress}h). Note that the continuum stress component will also contain contributions from diagonal and radial families of structural elements. 


\begin{figure}[!htbp]
    \centering
    \includegraphics[width=1.0\textwidth]{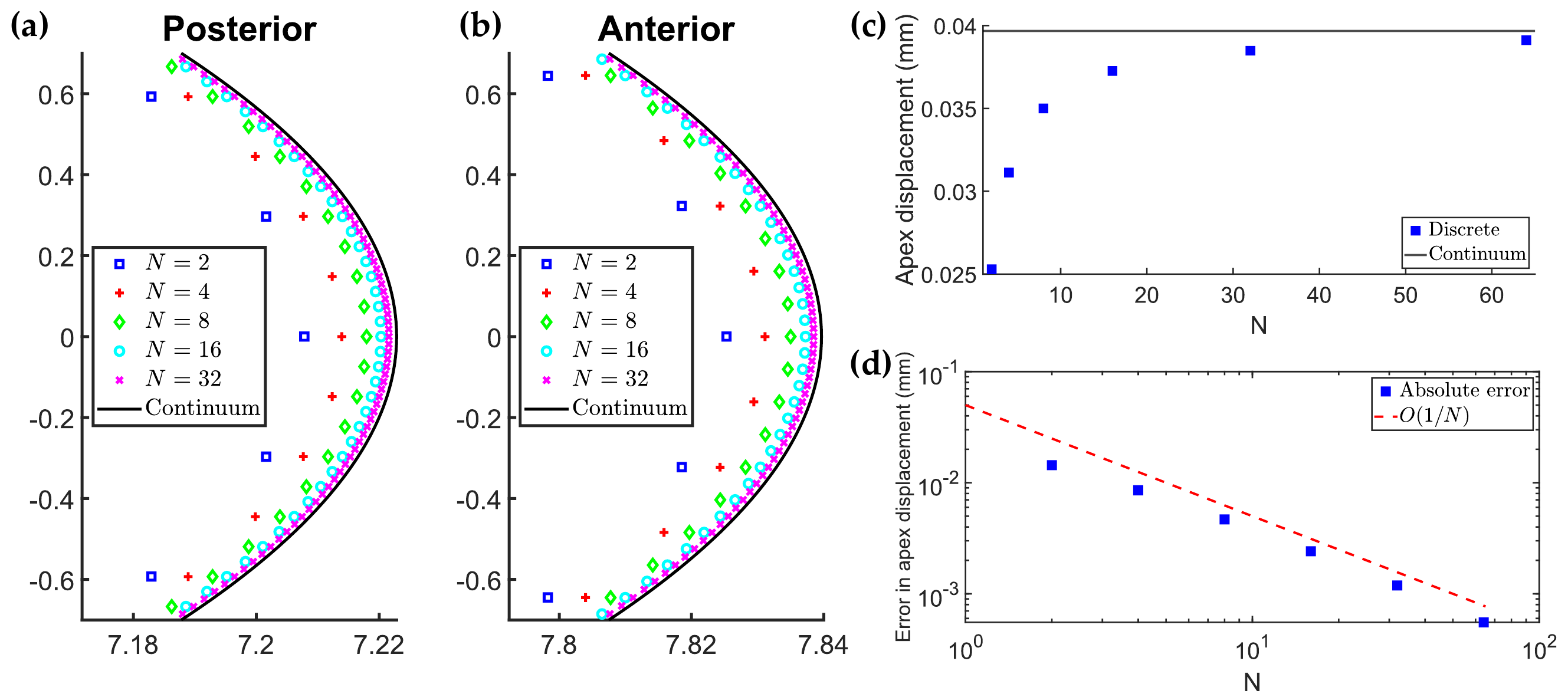}
    \caption{Details of posterior (a) and anterior (b) surfaces near the apex for a healthy cornea inflated to $2$ kPa IOP. Using $M=20N$, the discrete profiles for $N=2$ (blue squares), $N=4$ (red pluses), $N=8$ (green diamonds), $N=16$ (cyan circles) and $N=32$ (purple crosses) converge to the continuum predictions (solid black curves). Note that in order to make visible the otherwise small differences between the discrete and continuum solutions, we used different scales for $X$ and $Y$ axes (all indicated length units are still in mm). Panels on the right depict the predictions for apex displacement from discrete and continuum models. Panel (c) shows that the prediction of the discrete model for increasing $N$ (blue squares) approaches that of continuum model (horizontal black line). Panel (d) documents that the absolute error, defined as the difference in apex displacement between the discrete and the continuum model, decreases to $0$ as $O(1/N)$.}
    \label{fig:DiscreteToContinuumHealthyShape}
\end{figure}

Further details on the convergence of the predictions of the discrete model to those of the continuum is provided in Fig.~\ref{fig:DiscreteToContinuumHealthyShape}, which explores the convergence of the discrete-to-continuum upscaling, Fig.~\ref{fig:DiscreteToContinuumHealthyShape} compares the anterior and posterior profiles of the deformed cornea predicted by the discrete model with the corresponding profiles provided by the continuum model, as the number of lamellar layers increases. As expected, the predicted shapes of the discrete model approach the predicted shapes of the continuum model for growing $N$, on both posterior  (Fig.~\ref{fig:DiscreteToContinuumHealthyShape}a) and anterior surfaces (Fig.~\ref{fig:DiscreteToContinuumHealthyShape}b). For $N=32$ layers in the discrete model, the two profiles are almost indistinguishable.

Fig.~\ref{fig:DiscreteToContinuumHealthyShape}c shows the apex displacement in the radial direction for the discrete model as a function of the number of layers, showing that it gradually approaches the corresponding value predicted by the continuum model. The rate of this convergence is made explicit by plotting in Fig.~\ref{fig:DiscreteToContinuumHealthyShape}d  the absolute error between the predictions of the discrete and continuum models as a function of $N$; this absolute error scales as $N^{-1}$. 


The healthy cornea allows to establish a good qualitative and quantitative agreement between the discrete and continuum models. In the following, we proceed to explore the onset of keratoconus by using only the continuum model. 

\subsection{A damaged cornea: a model for keratoconus}
\label{ssec:keratoconus}

In order to simulate the degenerative eye disease keratoconus, we follow previous modelling work and impose a systematic reduction of the stiffness of the structure \cite{Pandolfi2019,Pandolfi2023}. In particular, we decrease the stiffness of both the lamellar segments ($\tilde{K}^{(1)}$) and the diagonal crosslinks ($\tilde{K}^{(3)}$) while holding the stiffness of the radial crosslinks  ($\tilde{K}^{(2)}$) fixed since their primary function is to keep collagen lamellae well-spaced. Since deterioration of microstructure is more pronounced near the corneal apex \cite{Pandolfi2006}, we impose a damage profile $\mathcal{D}(\tilde{Y})$ in the $Y$-direction which has a local maximum, specifically $\mathcal{D}_{max}$ ($0 \le \mathcal{D}_{max} \le 1$), at the centre of the cornea and is negligible at the limbus, as
\begin{equation}\label{eqn:DamageProfile}
\mathcal{D}(\tilde{Y}) = \mathcal{D}_{max} \left( 1- \left( \frac{\tilde{Y}}{\tilde{Y}_{max}} \right)^{\xi} \right),
\end{equation}
where $\tilde{Y}_{max}$ is the $\tilde{Y}$ coordinate of the node where the anterior surface meets the limbus. The spatial localisation of the damage profile can be modulated by varying the exponent $\xi=2,4,6,...$: profiles with larger $\xi$ have average damage closer to the maximum damage parameter. In the spirit of damage models, the damage profile affects the stiffness of the structural elements according to the linear relation
\begin{equation}
\overline{{\tilde K}^{(1)}}({\tilde Y}) = (1-{\mathcal D}){\tilde K}^{(1)}, \qquad \overline{{\tilde K}^{(3)}}({\tilde Y}) =(1-{\mathcal D}){\tilde K}^{(3)}.
\end{equation}

\begin{figure}[!htbp]
    \centering
    \includegraphics[width=0.8\textwidth]{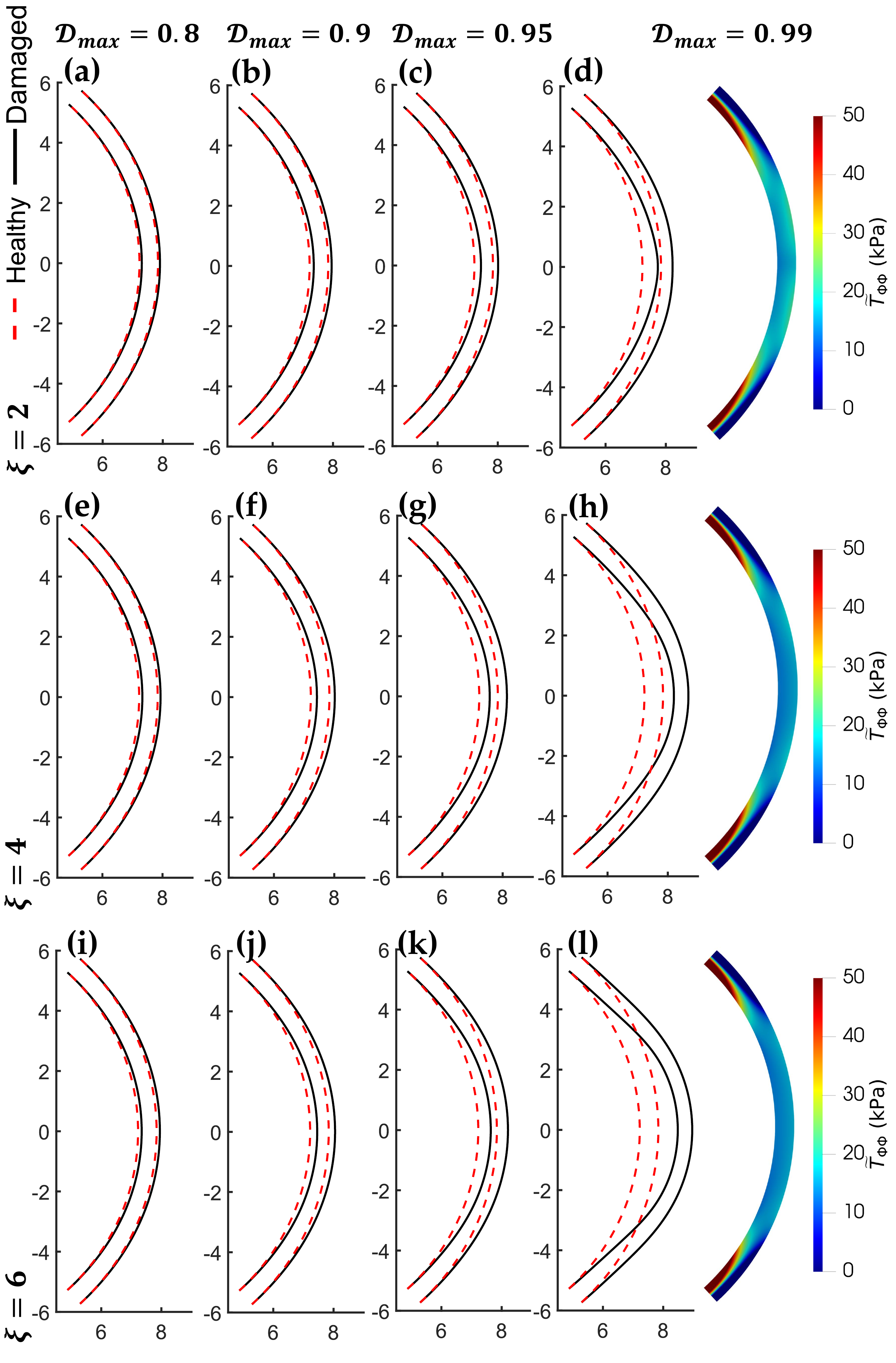}
    \caption{Predicted diseased corneal shapes using the continuum model with the parabolic ($\xi=2$ in \eqref{eqn:DamageProfile}; a-d), quartic ($\xi=4$; e-h) and sextic ($\xi=6$; i-l) damage profiles and varying value of central damage $\mathcal{D}_{max}=0.8$ (first column), $0.9$ (second column), $0.95$ (third column), $0.99$ (fourth column). Panels on the right show the circumferential stress $\tilde{T}_{\Phi \Phi}$ at $\mathcal{D}_{max} = 0.99$ for the three considered values of $\xi$ -- this can be compared with the healthy profile in Fig.~\ref{fig:DiscreteToContinuum_ForcesToStress}h.}
    \label{fig:VaryDamageParameters_Continuum}
\end{figure}

Fig.~\ref{fig:VaryDamageParameters_Continuum} shows the profiles of the anterior and posterior surfaces for several values of the maximum damage $\mathcal{D}_{max}$ and for several choices of the exponent $\xi$. The decrease in elemental stiffness in the damaged case results in more significant deformations compared to the healthy case, as expected. In particular, for a given value of the exponent $\xi$, the deformation becomes increasingly more pronounced as the maximum damage parameter is increased, where both the anterior and posterior surfaces bulge outwards, suggesting the formation of the conus. The profiles in Fig.~\ref{fig:VaryDamageParameters_Continuum}a-d are computed for $\xi=2$, in Fig.~\ref{fig:VaryDamageParameters_Continuum}e-h for $\xi=4$, and in Fig.~\ref{fig:VaryDamageParameters_Continuum}i-l for $\xi=6$. 
For a given value of the maximum damage parameter, the bulging of the anterior and posterior surfaces becomes more pronounced as the exponent $\xi$ increases.


The profile of the normal stress in the direction of the lamellae for the damaged cornea, Figs.~\ref{fig:VaryDamageParameters_Continuum}(d,h,l), is significantly different from the profile of the healthy case, where the approximately uniform stress near the apex is decreased but the large stress gradients adjacent to the limbus are greatly increased both in magnitude and in spatial extent, and visible over a much longer lengthscale compared to the healthy case. The amplification in the lamellar stress becomes even more pronounced as the spread of the damage is increased, $\xi$ is increased (Fig.~\ref{fig:VaryDamageParameters_Continuum}h,l). 


\begin{figure}[!htbp]
    \centering
    \includegraphics[width=1.0\textwidth]{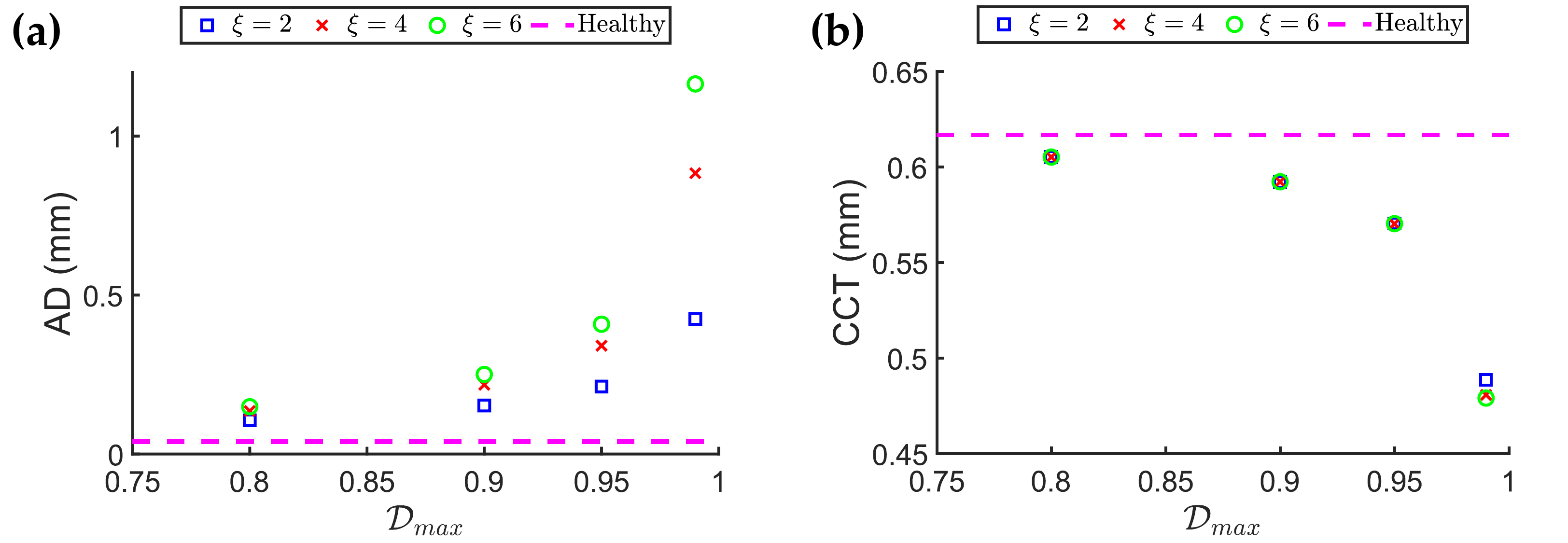}
    \caption{Key experimental metrics, apex displacement (AD; a) and central corneal thickness (CCT; b), plotted for varying damage parameters, $\mathcal{D}_{max}$ and $\xi$.}
    \label{fig:ContinuumKeyMetrics}
\end{figure}

To quantify the extent of corneal deformation following damage, Fig.~\ref{fig:ContinuumKeyMetrics}a plots the apex displacement versus the maximum damage parameter and Fig.~\ref{fig:ContinuumKeyMetrics}b plots the central corneal thickness  versus the maximum damage parameter. As expected from Fig.~\ref{fig:VaryDamageParameters_Continuum}, the apex displacement increases dramatically as the maximum damage parameter approaches one, becoming even more pronounced as the parameter $\xi$ increases. For the maximum damage parameter $\mathcal{D}_{max} = 0.99$, which is of the same order of magnitude as in previous studies \cite{Pandolfi2023}, the corresponding apex displacement exceeds $1$ mm. Furthermore, the corresponding thickness of the cornea decreases as a function of the maximum damage parameter, but the predictions are almost independent of the parameter $\xi$, Fig.~\ref{fig:ContinuumKeyMetrics}b. This can be explained by the fact that we do not reduce the stiffness of the radial crosslinks, which oppose the thinning of the cornea. We conclude that only the maximum damage parameter has a significant effect on the corneal thickness.



In what follows, we restrict attention to an extreme value of the maximum damage parameter, setting $\mathcal{D}_{max} = 0.99$, and fix the localisation parameter as $\xi=4$. 
\begin{figure}[!htbp]
    \centering
    \includegraphics[width=1.0\textwidth]{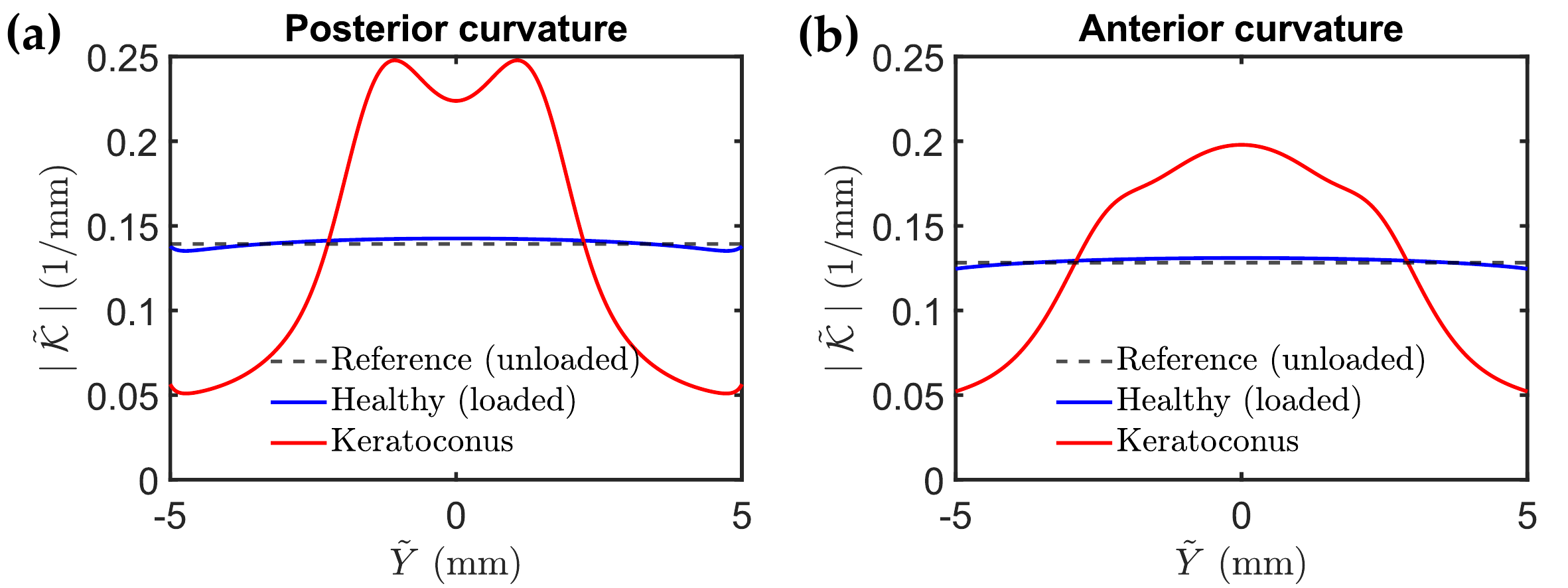}
    \caption{Comparison of reference, healthy and diseased (using $\mathcal{D}_{max}=0.99$ and $\xi=4$ in \eqref{eqn:DamageProfile}) curvatures. Panels (a) and (b) show curvatures of the posterior and anterior surface, respectively.}
    \label{fig:Damage99Curvatures}
\end{figure} 
In order to further quantify the change in corneal shape resulting from the imposed reduction in elemental stiffness, Fig.~\ref{fig:Damage99Curvatures} computes the curvature of both the posterior (Fig.~\ref{fig:Damage99Curvatures}a) and anterior surfaces (Fig.~\ref{fig:Damage99Curvatures}b) comparing the healthy (baseline) case to a heavily damaged case. We describe each surface in the current coordinates as $\tilde{x} = \tilde{F}(\tilde{y})$, from which we then compute the curvature \cite{Kreyszig2013Differential}
\begin{equation}
\label{eqn:CurvatureFormula}
\tilde{\mathcal{K}}(\tilde{F}) = \frac{\tilde{F}''}{\left( 1 + \left(\tilde{F}'\right)^{2} \right)^{3/2}}.
\end{equation}
In the healthy case, the curvature of the posterior and anterior surfaces are only mildly changed compared to the unloaded (constant) value: both surfaces have a local maximum (minimum) curvature at the centre-line, which gradually decreases toward the limbus (blue curves in Fig.~\ref{fig:Damage99Curvatures}). However, when the cornea is heavily damaged, the variations in surface curvature are much more significant (red curves in Fig.~\ref{fig:Damage99Curvatures}). The curvature profiles of both surfaces are in all cases symmetric about the centre-line, with anterior curvatures attaining a maximum at the centre $Y=0$. For $Y>0$ the curvature of the anterior surface decreases monotonically toward the limbus (Fig.~\ref{fig:Damage99Curvatures}b), but conversely the posterior curvature for a damaged cornea attains a local maximum for some finite value of $Y$ before decreasing (Fig.~\ref{fig:Damage99Curvatures}a). 


\begin{figure}[!htbp]
    \centering
    \includegraphics[width=0.95\textwidth]{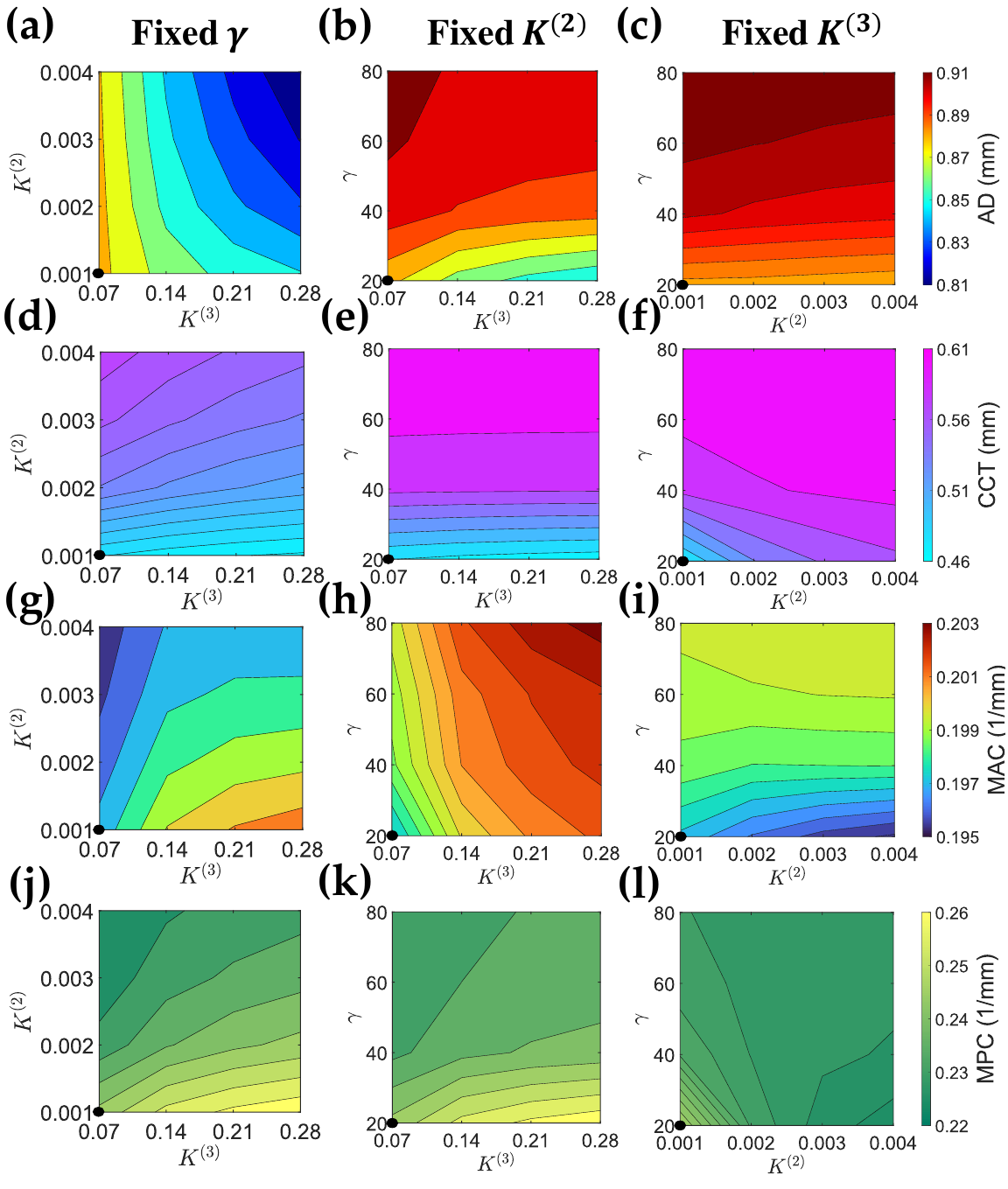}
    \caption{Comparison of key metrics of diseased corneas (using $\mathcal{D}_{max}=0.99$ and $\xi=4$ in \eqref{eqn:DamageProfile}) for varying $\gamma$, $K^{(2)}$ and $K^{(3)}$. In left, central and right panels we fix at their baseline values $\gamma$, $K^{(2)}$ and $K^{(3)}$ respectively and vary the remaining parameters. We plot the apex displacement (AD; a-c), the central corneal thickness (CCT; d-f), the maximum anterior curvature (MAC; g-i) and the maximum posterior curvature (MPC; j-k). For comparison, we note that the reference (unloaded cornea) values are AD=0 mm, CCT=0.62 mm, MAC =$1/\tilde{R}_A \approx 0.128$ mm$^{-1}$, MPC = $1/\tilde{R}_P \approx $ 0.139 mm$^{-1}$ and the values for healthy cornea loaded with physiological IOP are AD= 0.04 mm, CCT = 0.617 mm, MAC = 0.131 mm$^{-1}$ and MPC=0.143 mm$^{-1}$. Black circles at bottom-left corners of the panels indicate the case where all parameters are fixed at their baseline values ($\gamma=20$, $K^{(2)}=0.001$ and $K^{(3)}=0.07$).}
    \label{fig:VaryOtherParameters}
\end{figure}

We conclude with a sensitivity analysis of the continuum model predictions to the main parameters. Fig.~\ref{fig:VaryOtherParameters} presents a comprehensive sweep of the parameter spaces governed by the mesh aspect ratio parameter $\gamma$, the dimensionless stiffness of radial crosslinks $K^{(2)}$ and the dimensionless stiffness of the diagonal crosslinks $K^{(3)}$, illustrating their influence on the apex displacement (Fig.~\ref{fig:VaryOtherParameters}a-c), the central corneal thickness (Fig.~\ref{fig:VaryOtherParameters}d-f) and the maximum curvature of the anterior (Fig.~\ref{fig:VaryOtherParameters}g-i) and posterior surfaces (Fig.~\ref{fig:VaryOtherParameters}j-l). As expected, apex displacement is reduced as the stiffness of the radial and diagonal structural elements increases (Fig.~\ref{fig:VaryOtherParameters}a-c). The central corneal thickness is increased by increasing the stiffness of the radial elements (as these keep the lamellae spaced out, Fig.~\ref{fig:VaryOtherParameters}d,f) or by decreasing the stiffness of the diagonal crosslinks (which makes it easier to expand in the azimuthal direction, Fig.~\ref{fig:VaryOtherParameters}d,e). The dependency on the mesh aspect ratio parameter $\gamma$ is less transparent, as it is embedded within the definitions of the dimensional element stiffness (${\tilde K}^{(j)}$, $j=1,2,3$, which are proportional to $\gamma$) and the parameter $\kappa$ defined in Eq.~\eqref{eqn:DefsOf_q_kappa_w} (where increasing $\gamma$ essentially reduces the relative size of azimuthal derivatives compared to radial derivatives); see details in Supplementary Material. As a result, in simulations we observe that increasing the mesh aspect ratio parameter, $\gamma$, means that azimuthal stretching is more favorable than radial compression, resulting in a mild increase in the apex displacement as the cornea elongates (Fig.~\ref{fig:VaryOtherParameters}b,c) and a significant reduction in corneal thinning (Fig.~\ref{fig:VaryOtherParameters}e,f). The maximum curvature of the anterior surface is computed as the maximum value of Eq.~\eqref{eqn:CurvatureFormula}, excluding the outer 10\% at both ends to avoid the boundary effects at the rigidly pinned limbus. The maximum curvature of the posterior surface is calculated analogously. We identify only mild variation in the maximum anterior curvature across the parameter space, though for the baseline value of $\gamma$ it is mildly decreased by stiffening the radial crosslinks (Fig.~\ref{fig:VaryOtherParameters}g) and mildly increased by increasing the values of the mesh aspect ratio parameter and the stiffness of the diagonal crosslinks (Fig.~\ref{fig:VaryOtherParameters}g-i). The corresponding variation in the maximum posterior curvature is more evident, following the same trend as the anterior surface for changes in the stiffness of the radial and diagonal crosslinks (Fig.~\ref{fig:VaryOtherParameters}g-l), but for the baseline value of $K^{(2)}$ it is now very mildly decreased by increasing the values of the mesh aspect ratio parameter (Fig.~\ref{fig:VaryOtherParameters}k). 

\section{Discussion}
\label{sec:discussion}

In this paper we have developed a multiscale modelling framework for the mechanics of the human cornea when loaded with spatially and temporally uniform intraocular pressure, mimicking both the healthy shape with almost uniform curvature in baseline conditions (Fig.~\ref{fig:DiscreteToContinuum_ForcesToStress}), and a much more conical shape in response to a prescribed degeneration of the carrying structure (Fig.~\ref{fig:VaryDamageParameters_Continuum}). We proposed a discrete model composed of a planar regular grid of nodes interconnected by three families of structural elements (lamellar segments, radial and diagonal crosslinks, respectively), which contribute to a (plane-strain) elastic description of a thin corneal slice uniformly loaded with IOP (Figs.~\ref{fig:ReferenceGeometryDiscretized} and \ref{fig:DeformedGeometryDiscretized}). For simplicity, we assumed that all structural elements can be modelled as linear springs, and we estimated their corresponding elastic stiffness using measurements available in the literature. The simplicity of the discrete framework facilitated a rational upscaling of the discrete model into a continuum model (Eq.~\eqref{eqn:StrainEnergyLinearSprings}) which directly encodes the microscale properties of the structural elements, \emph{e.~g.} the stiffness of the individual structural elements, the spacing of the crosslinks. The predicted stresses in the collagen lamellae agree well between the discrete and continuum models, even for relatively coarse discretisations, including the boundary effects which arise due to pinning at the limbus, Fig.~\ref{fig:DiscreteToContinuum_ForcesToStress}. We further demonstrated that the predictions of the discrete model converge uniformly to the predictions of the continuum model as the number of collagen layers becomes large, Fig.~\ref{fig:DiscreteToContinuumHealthyShape}. Hence, the continuum model is a useful reduction of the discrete system which can be solved for a fraction of the computational cost, facilitating a wide survey of the parameter space. 

We employed this continuum model to explore how the key model parameters impact the macroscale corneal shape. We found that by systematically reducing the stiffness of the collagen lamealle and the diagonal crosslinks near the centre of cornea, in an attempt to mimick damage to the carrying structure, the cornea became outwardly bulged with a significant reduction in thickness (Fig.~\ref{fig:VaryDamageParameters_Continuum}). The precise shapes of the corneal interfaces were only weakly dependent on the spatial profile of the applied damage (Fig.~\ref{fig:VaryDamageParameters_Continuum}). For highly damaged profiles, where the stiffness of structural elements at the centre of the cornea was set to be only 1\% of the baseline value, we observed formation of a conical corneal shape and significant thinning (Figs.~\ref{fig:VaryDamageParameters_Continuum}, \ref{fig:ContinuumKeyMetrics}), typical of corneal shapes observed in patients suffering from the degenerative keratoconus. Although this level of damage is extreme, this value is in line with those applied in other models in order to see large scale deformation of the cornea \cite{Pandolfi2019,Pandolfi2023}. Furthermore, the observation that a large amount of damage is required to see significant deformation of the structure is consistent with the known stiffening of truss-work structures induced by diagonal elements \cite{Pandolfi2019}.

Although two-dimensional geometry considered in this study is highly idealized, it provides a rational framework that can readily be adapted to include more physical ingredients. For example, although the current model neglects the complex three-dimensional arrangement of collagen in a human cornea \cite{Meek2009}, adding a third spatial dimension to the discrete model should not significantly increase the complexity of the governing equations, although it will increase the computational run time. The upscaling to a three-dimensional continuum model should work analogously. Similarly, the discrete and continuum models could easily accommodate gradual stiffness changes across the structural elements, \emph{e.~g.} from the posterior surface to the anterior \cite{Petsche2012}, or from limbus to limbus due to variations in the distribution of microfibril bundles \cite{Alberts2017,White2017Elastic}. Furthermore, the discrete model could also be adapted to consider additional complexities such as through-thickness variation in the out-of-plane lamellar inclination, lamellar interweaving and inhomogeneous distribution of preferential in-plane lamellar orientations, although in these cases the upscaling to a continuum model could become more challenging.

One major advantage of the discrete framework proposed herein is the point-wise control over geometrical and material properties of the carrying structure. Hence, this discrete model provides an ideal framework for investigation of the role of ECM composition in promoting changes in corneal shape \cite{Patey1984,Pouliquen1987,Fullwood1992,Wollensak1990,Sawaguchi1991}, as well as informing protocols for corneal crosslinking therapies designed to arrest progression of keratoconus \cite{Raiskup2008Collagen,Baiocchi2009Corneal}. The continuum model could similarly be adapted to include spatially dependent geometric and material properties. However, investigation of these questions is postponed to future work.

\vskip6pt

\enlargethispage{20pt}


\providecommand{\dataaccess}[1]{\textbf{\textit{Data accessibility: }} #1}

\dataaccess{This article has no experimental data. Numerical scripts for the discrete model were written in Matlab version R2021a, those solving the continuum model were written in python using FEniCS version 2019.2.0.dev0--, and can be accessed at \url{http://dx.doi.org/10.5525/gla.researchdata.1548}.}

\providecommand{\funding}[1]{\textbf{\textit{Funding: }} #1}

\funding{J.K., N.A.H., X.Y.L. and P.S.S. acknowledge funding from EPSRC grant no. EP/S030875/1.}

\providecommand{\ack}[1]{\textbf{\textit{Acknowledgements: }} #1}

\ack{A.P. acknowledges the Italian National Group of Physics-Mathematics (GNFM) of the Italian National Institution of High Mathematics ‘‘Francesco Severi’’ (INDAM).}


\bibliography{references} 
\bibliographystyle{plain} 

\makeatletter\@input{xxSupp.tex}\makeatother

\end{document}


\title{A discrete-to-continuum model for the human cornea with application to keratoconus \\ \vspace{0.5cm}
\Large Supplementary Material}


\author{
J. Köry$^{1}$, P.~S. Stewart$^{1}$, N.~A. Hill$^{1}$,
X.~Y. Luo$^{1}$ and A. Pandolfi$^{2}$}

\affil{$^{1}$School of Mathematics and Statistics, University of Glasgow,  Glasgow, G12 8QQ, U.K.\\
$^{2}$Department of Civil and Environmental Engineering, Politecnico di Milano, Piazza Leonardo da Vinci 32, 20133 Milano, Italy}

\maketitle



This Supplementary Material contains two sections. Sec. \ref{sec:EstimateStiffnessParameters} explains how we estimated stiffness parameters in our model. Sec. \ref{app:upscaling} contains within its subsections detailed derivations pertaining to discrete-to-continuum upscaling (including that of the posterior boundary condition), transformations between the coordinate systems as well as between different configurations, derived continuum-level quantities including the strain energy density and stress tensors and how these relate to their small-deformations limit (linear elasticity).

\section{Estimation of stiffness parameters}
\label{sec:EstimateStiffnessParameters}
\subsection{Collagen lamellae}
\label{ssec:collagenLamellae}

A material model for a single collagen lamella, described as a matrix (composed of proteoglycans, water and salt ions) reinforced with parallel arrays of collagen fibrils, was proposed first in \cite{Petsche2013Supp}. By using detailed information on the variation of the mean fibril direction across the thickness from second harmonic-generated imaging, the authors built a finite element model used to simulate \emph{ex vivo} experimental tests on cornea buttons. The simulations were used to identify the material parameters of the model, among which the ones of the lamellae. The collagen fibrils were assumed to be nonlinearly elastic, characterized by a strain energy density of exponential type \cite{gasser2006Supp}, depending on the square of the stretch in the fibrils  direction. In the special case that the fibrils are all oriented parallel to the longitudinal axis of the lamella, the strain energy density assumes the simple form
$$\tilde{w}^{(1)}(\lambda) = \frac{\tilde{\alpha}_1}{2 \alpha_2} 
\left \{ \exp \left[ {\alpha_2 (\lambda^2 - 1)^2 } \right] - 1 \right \} \, .$$
If the deformation is uniform over the cross-section of the lamella, we can define a strain energy per unit of length as
$$\tilde{W}^{(1)}(\lambda) = \tilde{A} \tilde{w}^{(1)}(\lambda)\, ,$$
The force magnitude $\tilde{f}^{(1)}(\lambda)$ follows by differentiation with respect to the stretch $\lambda$
\begin{equation}\label{eqn:TensileForceLamellaDiscrete}
    \tilde{f}^{(1)}(\lambda) 
    =
    \frac{d \tilde{W}^{(1)}}{d \lambda} 
    =
    2 \tilde{\alpha}_1 \tilde{A} \lambda \left( \lambda^2  -1 \right) e^{\alpha_2 \left( \lambda^2 -1  \right)^2} \, .
\end{equation}
The linear behavior derives by assuming small strains, i.~e., $\lambda = 1 + \varepsilon$, with $\varepsilon \ll 1$. By neglecting higher-order terms we obtain
\begin{equation}\label{eqn:linearization}
    \tilde{f}^{(1)}(\lambda)
    \approx 
    4 \tilde{\alpha}_1 \tilde{A} \varepsilon 
    = 
    4 \tilde{\alpha}_1 \tilde{A} (\lambda - 1)
    = 
    \frac{4 \tilde{\alpha}_1 \tilde{A} }{\tilde{L}} \left(\tilde{l} - \tilde{L}\right) \, .
\end{equation}
A comparison with Eq.~\eqref{eqn:LinearSpringsGeneralModel} indicates that, within the linear approximation, the stiffness of the element representative of the lamellae is
\begin{equation}\label{eqn:LinearStiffnessYoungModulusFromPetschePinsky}
    \tilde{K}_{\rm sph}^{(1)} = \frac{4 \tilde{\alpha}_1 \tilde{A}}{\tilde{L}} \, ,
\end{equation}
where the term $4 \tilde{\alpha}_1$ plays the role of the elastic modulus $\tilde{E}$ in the standard stiffness of a element and we use subscript sph to indicate that this estimate of stiffness applies to real three-dimensional corneal geometry that is approximated as spherical (as opposed to the two-dimensional geometry used in this work). Using the values obtained from an identification procedure reported in  \cite{Petsche2013Supp}, i.~e., $\tilde{\alpha}_1 = 638$~kPa, we estimate the equivalent elastic modulus for the lamellae as $\tilde{E}_{\rm sph}^{(1)} = 4 \tilde{\alpha}_1 = 2 552$ kPa.

We next estimate the appropriate stiffness for lamellar segments in our two-dimensional setting. This setting corresponds to considering a meridian slice of the cornea and thus describes a cylindrical structure uniformly loaded with IOP so that we can reduce the problem to a plane-strain description (below we will provide a correction accounting for this fact). We can estimate lamellar area $\tilde{A}$ as the unloaded corneal thickness $\tilde{T}$ divided by the number of subdivisions $N$ times the depth in the out-of-plane direction $\tilde{D}$, while the lamellar length $\tilde{L}$ can be estimated as the length of the meridian $ 2\tilde{R}_A \Phi^*$ divided by the number of subdivisions $M$ (for large $M$), obtaining a representative value of the lamellar stiffness $\tilde{K}_{\rm sph}^{(1)}$ as
\begin{equation}\label{eqn:LinearSpringsGeneralModel2}
    \tilde{K}_{\rm sph}^{(1)} = \tilde{E}_{\rm sph}^{(1)} \tilde{D} \frac{\tilde{T}}{N} \frac{M}{ 2\tilde{R}_A \Phi^*} = 
    \gamma \frac{\tilde{E}_{\rm sph}^{(1)} \tilde{D} \tilde{T}}{ 2\tilde{R}_A \Phi^*} \, , 
\end{equation}
which, using the $\gamma=20$ adopted in the numerical calculations and $\tilde{D} = 1$ mm, provides $\tilde{K}_{\rm sph}^{(1)} \approx 2.4 \text{ N/mm}$. 
As already mentioned, our two-dimensional model setting needs to be interpreted as a cylindrically-shaped shell under plane strain which means that we have not yet accounted for the three-dimensional shape of a sphere. It is well known that the mechanical response of a cylindrical shell is less stiff than the one of a spherical shell, thus the stiffness of the components of our model should be suitably increased. 
An approximation of the increment to assign to the stiffness of the elements can be obtained from the Laplace's law applied to a sphere and a cylinder respectively.
For an isotropic thin membrane of thickness $\tilde{T}$, Young's modulus $\tilde{E}$, Poisson's ratio $\nu$, and unloaded radius of curvature $\tilde{R}_A$ relates the circumferential strain $\varepsilon_{\theta}$ to the internal pressure $\tilde{p}$ as
$$\tilde{p} = \frac{2 \tilde{T} \tilde{E}_{\rm sph} \varepsilon_{\theta}}{\tilde{R}_A \left( 1 - \nu \right)} 
\qquad \text{for a sphere and} 
\qquad \tilde{p} = \frac{2 \tilde{T} \tilde{E}_{\rm cyl} \varepsilon_{\theta}}{\tilde{R}_A \left( 2 - \nu \right)} \qquad \text{for a cylinder}.$$
The two geometries give the same hoop strain under the same pressure if the two elastic moduli are related as
$$ \tilde{E}_{\rm cyl} = \frac{2- \nu}{1-\nu} \tilde{E}_{\rm sph} \, .$$
Biological materials are typically characterized by a low compressibility, which for a linear elastic isotropic material requires $\nu \approx 0.5$. In this work we will thus use the adjusted value $\tilde{E}^{(1)} = \tilde{E}_{\rm cyl}^{(1)} = 3 \tilde{E}_{\rm sph}^{(1)} \approx 7,656$ kPa.
Following this suggestion, the stiffness of the two-dimensional model is modified as 
$$\tilde{K}_{\rm cyl}^{(1)}  = \frac{2 - \nu}{1- \nu} \tilde{K}_{\rm sph}^{(1)} ,$$
leading to an adjusted value $\tilde{K}^{(1)} = 3 \tilde{K}_{\rm sph}^{(1)} \approx 7.2$ N/mm.

Finally note that it is possible for the lamellar segments to exhibit distinct stiffnesses in response to tension and compression. However, due to the absence of reliable experimental estimates, we simply assume that these two stiffnesses are equal.

\subsection{Diagonal and radial cross-links}
\label{ssec:CLSMatrix}

In the lack of information concerning the mechanical properties of the underlying ground matrix, for the radial and diagonal cross-links we for simplicity assume  a linear behavior characterized by two stiffness parameters, i.~e., $\tilde{K}^{(2)}$ for the radial cross-links and $\tilde{K}^{(3)}$ for the diagonal cross-links. Since the collagen lamellae are the carrying components of the cornea, it makes sense to assign the two parameters smaller values than the one of the lamellar stiffness $\tilde{K}^{(1)}$. In Sec. \ref{sec:upscaling}, we upscale the discrete model into continuum by increasing $N$ to infinity whilst keeping $\gamma$ constant. In our geometrical model, the lengths and the equivalent areas of the radial and diagonal cross-links thus reduce according to the reduction of lengths and areas of the collagen lamellae. Therefore we simply assume that, in all the discretizations, the ratio between radial and diagonal cross-link stiffness parameters and lamellar stiffness remain constant.\footnote{Let us further note that while the lengths of individual elements $\tilde{L}$ scale as $1/N$, by requiring that the total volume (mass) of the network be conserved, one can infer the same scaling for the cross-sectional areas $\tilde{A}$. Thus, it follows that the stiffnesses $\tilde{K}^{(m)}$ of all elements from \eqref{eqn:LinearSpringsGeneralModel} remain constant as $N$ (as well as $M =\gamma N$) is increased.} The stiffness ratios of these components can be derived from nano-scale experiments on cross-links (see the discussion section in \cite{Pandolfi2019Supp}). Here we impose as baseline the ratios from \cite{Pandolfi2023Supp} and assume $\tilde{K}^{(2)} = 0.001~\tilde{K}^{(1)} $ and $\tilde{K}^{(3)} = 0.07~\tilde{K}^{(1)}$. While both radial and diagonal cross-links represent the ground matrix (see Sec.~\ref{ssec:discretization}), radial cross-links will be under compression due to the action of IOP and the diagonal cross-links will (typically) be under tension. Experimental work from \cite{Haverkamp2005Supp} provides some evidence that the stiffness of proteoglycans in tension (family 3 in our discrete model) can indeed be much greater than that in compression (family 2).


\section{Upscaling and the continuum model}
\label{app:upscaling}

\subsection{Discrete-to-continuum upscaling}

\begin{adjustwidth}{-30pt}{0pt}

We will use polar $R$ and $\Phi$ as independent variables and search for equations to be satisfied by the continuum functions $x(R,\Phi)$ and $y(R,\Phi)$ for given $q(R)$. It is useful to introduce the following operators
\begin{equation}\label{eqn:OperatorsDefinedPolarNew}
A_- = - \frac{\partial}{\partial R} \qquad  A_+ = \frac{\partial}{\partial R} \qquad  B_- = - \kappa \frac{\partial}{\partial \Phi} \qquad B_+ = \kappa \frac{\partial}{\partial \Phi}.   
\end{equation}
Assuming the separation of length scales, i.e. $\ve \ll 1$, Taylor expansions of the finite differences then yield
$$ x_{i,j-1}  - x_{i,j} = \ve B_-(x) + \frac{\ve^2}{2} B_-^2(x) + O(\ve^3 ) \qquad x_{i,j+1}  - x_{i,j} = \ve B_+(x) + \frac{\ve^2}{2} B_+^2(x) + O(\ve^3 ) $$
$$ x_{i-1,j}  - x_{i,j} = \ve A_-(x) + \frac{\ve^2}{2} A_-^2(x) + O(\ve^3 ) \qquad x_{i+1,j}  - x_{i,j} = \ve A_+(x) + \frac{\ve^2}{2} A_+^2(x) + O(\ve^3 ) $$
$$ x_{i-1,j-1}  - x_{i,j} = \ve \left(A_- + B_-\right) (x) + \frac{\ve^2}{2} \left(A_- + B_-\right)^2 (x) + O(\ve^3 ) \qquad x_{i-1,j+1}  - x_{i,j} = \ve \left(A_- + B_+\right) (x) + \frac{\ve^2}{2} \left(A_- + B_+\right)^2 (x) + O(\ve^3 ) $$
$$ x_{i+1,j-1}  - x_{i,j} = \ve \left(A_+ + B_-\right) (x) + \frac{\ve^2}{2} \left(A_+ + B_-\right)^2 (x) + O(\ve^3 ) \qquad x_{i+1,j+1}  - x_{i,j} = \ve \left(A_+ + B_+\right) (x) + \frac{\ve^2}{2} \left(A_+ + B_+\right)^2 (x) + O(\ve^3 ) $$
and same for $y(X,Y)$. Furthermore, we get
\begin{equation}\label{eqn:DefOfMu}
w_{i \pm \frac12} = \frac{\sqrt{q_i q_{i \pm 1}}}{\sqrt{1+q_i q_{i \pm 1}}} = \frac{\sqrt{q (q + \ve A_{\pm}(q) + O(\ve^2))}}{\sqrt{1 + q (q + \ve A_{\pm}(q) + O(\ve^2))}} = \frac{q}{\sqrt{1+q^2}} + \frac{\ve}{2} A_{\pm} \left( \frac{q}{\sqrt{1+q^2}} \right) + O(\ve^2).
\end{equation}
(Dividing by $\varepsilon$) The force balance \eqref{eqn:dimLessForceBalanceCircularNew} thus becomes
$$ 0 = g^{(1)} \left( q \sqrt{ \left( B_-(x) \right)^2 + \left( B_-(y) \right)^2 + \ve \left( B_-(x) B_-^2(x) + B_-(y) B_-^2(y) \right) + O(\ve^2)  } \right) \times$$
$$ \left( B_-(x) + \frac{\ve}{2} B_-^2(x) + O(\ve^2), B_-(y) + \frac{\ve}{2} B_-^2(y) + O(\ve^2)\right) + $$
$$g^{(1)}\left( q \sqrt{ \left( B_+(x) \right)^2 + \left( B_+(y) \right)^2 + \ve \left( B_+(x) B_+^2(x) + B_+(y) B_+^2(y) \right) + O(\ve^2)  } \right) \times$$
$$ \left( B_+(x) + \frac{\ve}{2} B_+^2(x) + O(\ve^2), B_+(y) + \frac{\ve}{2} B_+^2(y) + O(\ve^2)\right) + $$
$$ g^{(2)} \left( \sqrt{ \left( A_-(x) \right)^2 + \left( A_-(y) \right)^2 + \ve \left( A_-(x) A_-^2(x) + A_-(y) A_-^2(y) \right) + O(\ve^2)  } \right) \times$$
$$ \left( A_-(x) + \frac{\ve}{2} A_-^2(x) + O(\ve^2), A_-(y) + \frac{\ve}{2} A_-^2(y) + O(\ve^2)\right) + $$
$$ g^{(2)}\left( \sqrt{ \left( A_+(x) \right)^2 + \left( A_+(y) \right)^2 + \ve \left( A_+(x) A_+^2(x) + A_+(y) A_+^2(y) \right) + O(\ve^2)  } \right) \times$$
$$ \left( A_+(x) + \frac{\ve}{2} A_+^2(x) + O(\ve^2), A_+(y) + \frac{\ve}{2} A_+^2(y) + O(\ve^2)\right) + $$
$$ g^{(3)} \left( \left(\frac{q}{\sqrt{1+q^2}} + \frac{\ve}{2} A_{-} \left( \frac{q}{\sqrt{1+q^2}}\right) +O(\ve^2) \right) \times \right. $$
$$ \left. \sqrt{ \left( (A_- + B_-)(x) \right)^2 + \left(\left(A_- + B_-\right)(y) \right)^2 + \ve \left( ((A_- + B_-)(x))((A_- + B_-)^2(x)) +  ((A_- + B_-)(y))((A_- + B_-)^2(y)) \right) + O(\ve^2)  } \right) \times$$
$$ \left( (A_- + B_-)(x) + \frac{\ve}{2} (A_- + B_-)^2(x)  + O(\ve^2), (A_- + B_-)(y) + \frac{\ve}{2} (A_- + B_-)^2(y)  + O(\ve^2) \right) + $$
$$ g^{(3)} \left(  \left(\frac{q}{\sqrt{1+q^2}} + \frac{\ve}{2} A_{-} \left( \frac{q}{\sqrt{1+q^2}}\right) +O(\ve^2) \right) \times \right.$$
$$\left. \sqrt{ \left( (A_- + B_+)(x) \right)^2 + \left(\left(A_- + B_+\right)(y) \right)^2 + \ve \left( ((A_- + B_+)(x))((A_- + B_+)^2(x)) +  ((A_- + B_+)(y))((A_- + B_+)^2(y)) \right) + O(\ve^2)  } \right) \times$$
$$\left( (A_- + B_+)(x) + \frac{\ve}{2} (A_- + B_+)^2(x)  + O(\ve^2), (A_- + B_+)(y) + \frac{\ve}{2} (A_- + B_+)^2(y)  + O(\ve^2) \right) + $$
$$ g^{(3)} \left( \left(\frac{q}{\sqrt{1+q^2}} + \frac{\ve}{2} A_{+} \left( \frac{q}{\sqrt{1+q^2}}\right) +O(\ve^2) \right) \times \right. $$
$$ \left. \sqrt{ \left( (A_+ + B_-)(x) \right)^2 + \left(\left(A_+ + B_-\right)(y) \right)^2 + \ve \left( ((A_+ + B_-)(x))((A_+ + B_-)^2(x)) +  ((A_+ + B_-)(y))((A_+ + B_-)^2(y)) \right) + O(\ve^2)  } \right) \times$$
$$ \left( (A_+ + B_-)(x) + \frac{\ve}{2} (A_+ + B_-)^2(x)  + O(\ve^2), (A_+ + B_-)(y) + \frac{\ve}{2} (A_+ + B_-)^2(y)  + O(\ve^2) \right) + $$
$$ g^{(3)} \left( \left(\frac{q}{\sqrt{1+q^2}} + \frac{\ve}{2} A_{+} \left( \frac{q}{\sqrt{1+q^2}}\right) +O(\ve^2) \right) \times \right. $$
$$ \left. \sqrt{ \left( (A_+ + B_+)(x) \right)^2 + \left(\left(A_+ + B_+\right)(y) \right)^2 + \ve \left( ((A_+ + B_+)(x))((A_+ + B_+)^2(x)) +  ((A_+ + B_+)(y))((A_+ + B_+)^2(y)) \right) + O(\ve^2)  } \right) \times$$
$$ \left( (A_+ + B_+)(x) + \frac{\ve}{2} (A_+ + B_+)^2(x)  + O(\ve^2), (A_+ + B_+)(y) + \frac{\ve}{2} (A_+ + B_+)^2(y)  + O(\ve^2) \right),$$
where we (from now onwards) employ notation $(v_1,v_2)$ for Cartesian coordinates of a vector $\boldsymbol{v}$, i.e.
$$\boldsymbol{v} = v_1 \hat{\boldsymbol{x}} + v_2 \hat{\boldsymbol{y}}.$$
We Taylor expand everything assuming $\ve \ll 1$. Noting that $B_- = - B_+$ and $A_- = - A_+$, the balance at $O(1)$ is trivially satisfied and at $O(\ve)$ we get
$$ 0 = B_+ \left( g^{(1)}\left( q \sqrt{(B_+(x))^2+(B_+(y))^2} \right) \left(B_+(x),B_+(y) \right) \right) + A_+ \left( g^{(2)}\left(\sqrt{(A_+(x))^2+(A_+(y))^2} \right) \left(A_+(x),A_+(y) \right) \right) +$$
$$ \left( A_+ + B_+ \right) \left( g^{(3)}\left( \frac{q}{\sqrt{1+q^2}} \sqrt{((A_+ + B_+)(x))^2+((A_+ + B_+)(y))^2} \right) \left((A_+ + B_+)(x),(A_+ + B_+)(y) \right) \right) +$$
$$ \left( A_+ + B_- \right) \left( g^{(3)}\left( \frac{q}{\sqrt{1+q^2}} \sqrt{((A_+ + B_-)(x))^2+((A_+ + B_-)(y))^2} \right) \left((A_+ + B_-)(x),(A_+ + B_-)(y) \right) \right).$$
This can be rewritten recalling the definitions of $A_+$, $B_+$, $B_-$ and $w(R)$ as Eq. \eqref{eqn:UpscaledProblem1}.

\subsection{Transforming dependent variables from Cartesian to polar  coordinates}

Next, we transform the (dependent) Cartesian variables into polar as
$$ x(R,\Phi) = r(R,\Phi) \cos{\phi(R,\Phi)} \qquad y(R,\Phi) = r(R,\Phi) \sin{\phi(R,\Phi)}$$
and get
$$ x_R = r_R \cos{\phi} - r \sin{\phi} \phi_R \quad x_\Phi = r_\Phi \cos{\phi} - r \sin{\phi} \phi_\Phi \quad y_R = r_R \sin{\phi} + r \cos{\phi} \phi_R \quad y_\Phi = r_\Phi \sin{\phi} + r \cos{\phi} \phi_\Phi $$
from which it follows that
$$ x_R^2 + y_R^2 = r_R^2 + \left(r \phi_R \right)^2 \quad x_\Phi^2 + y_\Phi^2 = r_\Phi^2 + \left(r \phi_\Phi \right)^2 \quad \left( x_R \pm \kappa x_\Phi \right)^2 + \left( y_R \pm \kappa y_\Phi \right)^2 = \left( r_R \pm \kappa r_\Phi \right)^2 + \left( r \Phi_R \pm r \kappa \phi_\Phi \right)^2 $$
and
$$\frac{q}{\sqrt{1+q^2}} =  \frac{1}{\sqrt{1+\kappa^2 R^2}}.$$
This results in
$$ 0= \left( g^{(2)} \left( \sqrt{r_R^2 + (r \phi_R)^2} \right) \left(r_R \cos{\phi} - r \sin{\phi} \phi_R,r_R \sin{\phi} + r \cos{\phi} \phi_R \right) + g^{(3)} \left( \sqrt{\frac{\left(  r_R + \kappa r_\Phi \right)^2 + \left( r \phi_R + r \kappa \phi_\Phi \right)^2 }{1+ \kappa^2 R^2} } \right)  \times \right. $$
$$ \left(  r_R \cos{\phi} - r \sin{\phi} \phi_R + \kappa (r_\Phi \cos{\phi} - r \sin{\phi} \phi_\Phi), r_R \sin{\phi} + r \cos{\phi} \phi_R + \kappa (r_\Phi \sin{\phi} + r \cos{\phi} \phi_\Phi) \right)  + $$
$$ \left. g^{(3)} \left( \sqrt{\frac{\left(  r_R - \kappa r_\Phi \right)^2 + \left( r \phi_R - r \kappa \phi_\Phi \right)^2 }{1+ \kappa^2 R^2} } \right)  \times \right. $$
$$ \left. \left(  r_R \cos{\phi} - r \sin{\phi} \phi_R - \kappa (r_\Phi \cos{\phi} - r \sin{\phi} \phi_\Phi), r_R \sin{\phi} + r \cos{\phi} \phi_R - \kappa (r_\Phi \sin{\phi} + r \cos{\phi} \phi_\Phi) \right) \right)_R +
 $$
$$ \kappa \left( \kappa  g^{(1)} \left( \frac{\sqrt{r_\Phi^2 + (r \phi_\Phi)^2}}{R} \right) \left(r_\Phi \cos{\phi} - r \sin{\phi} \phi_\Phi, r_\Phi \sin{\phi} + r \cos{\phi} \phi_\Phi \right) +  g^{(3)} \left( \sqrt{\frac{\left(  r_R + \kappa r_\Phi \right)^2 + \left( r \phi_R + r \kappa \phi_\Phi \right)^2 }{1+ \kappa^2 R^2} } \right)  \times \right. $$
$$ \left(  r_R \cos{\phi} - r \sin{\phi} \phi_R + \kappa (r_\Phi \cos{\phi} - r \sin{\phi} \phi_\Phi), r_R \sin{\phi} + r \cos{\phi} \phi_R + \kappa (r_\Phi \sin{\phi} + r \cos{\phi} \phi_\Phi) \right)  - $$
$$ \left. g^{(3)} \left( \sqrt{\frac{\left(  r_R - \kappa r_\Phi \right)^2 + \left( r \phi_R - r \kappa \phi_\Phi \right)^2 }{1+ \kappa^2 R^2} } \right)  \times \right. $$
$$ \left. \left(  r_R \cos{\phi} - r \sin{\phi} \phi_R - \kappa (r_\Phi \cos{\phi} - r \sin{\phi} \phi_\Phi), r_R \sin{\phi} + r \cos{\phi} \phi_R - \kappa (r_\Phi \sin{\phi} + r \cos{\phi} \phi_\Phi) \right) \right)_\Phi$$

\subsection{Transforming unit vectors from Cartesian to polar coordinates}

It is easy to show that upon transforming the unit vectors from cartesian to polar coordinates
$$ \hat{\boldsymbol{x}} = \cos{\phi} \hat{\boldsymbol{r}} - \sin{\phi} \hat{\boldsymbol{\phi}} \qquad \hat{\boldsymbol{y}} = \sin{\phi} \hat{\boldsymbol{r}} + \cos{\phi} \hat{\boldsymbol{\phi}}, $$
we get
$$ 0= \left( g^{(2)} \left( \sqrt{r_R^2 + (r \phi_R)^2} \right) \left(r_R \hat{\boldsymbol{r}} + r \phi_R \hat{\boldsymbol{\phi}} \right) + g^{(3)} \left( \sqrt{\frac{\left(  r_R + \kappa r_\Phi \right)^2 + \left( r \phi_R + r \kappa \phi_\Phi \right)^2 }{1+ \kappa^2 R^2} } \right) \left((r_R + \kappa r_\Phi) \hat{\boldsymbol{r}} + (r \phi_R + \kappa r \phi_\Phi) \hat{\boldsymbol{\phi}}\right)  + \right. $$
$$ \left. g^{(3)} \left( \sqrt{\frac{\left(  r_R - \kappa r_\Phi \right)^2 + \left( r \phi_R - r \kappa \phi_\Phi \right)^2 }{1+ \kappa^2 R^2} } \right) \left( (r_R - \kappa r_\Phi) \hat{\boldsymbol{r}} + (r \phi_R - \kappa r \phi_\Phi) \hat{\boldsymbol{\phi}} \right) \right)_R + $$
$$ \kappa \left( \kappa  g^{(1)} \left( \frac{\sqrt{r_\Phi^2 + (r \phi_\Phi)^2}}{R} \right) \left( r_\Phi \hat{\boldsymbol{r}} + r \phi_\Phi \hat{\boldsymbol{\phi}} \right) +   g^{(3)} \left( \sqrt{\frac{\left(  r_R + \kappa r_\Phi \right)^2 + \left( r \phi_R + r \kappa \phi_\Phi \right)^2 }{1+ \kappa^2 R^2} } \right) \left((r_R + \kappa r_\Phi) \hat{\boldsymbol{r}} + (r \phi_R + \kappa r \phi_\Phi) \hat{\boldsymbol{\phi}} \right)  - \right. $$
$$ \left. g^{(3)} \left( \sqrt{\frac{\left(  r_R - \kappa r_\Phi \right)^2 + \left( r \phi_R - r \kappa \phi_\Phi \right)^2 }{1+ \kappa^2 R^2} } \right) \left((r_R - \kappa r_\Phi) \hat{\boldsymbol{r}} + (r \phi_R - \kappa r \phi_\Phi) \hat{\boldsymbol{\phi}} \right) \right)_\Phi $$
In the current configuration one has
\begin{equation}
\frac{\partial \hat{\boldsymbol{r}}}{\partial \phi} = \hat{\boldsymbol{\phi}} \qquad \frac{\partial \hat{\boldsymbol{\phi}}}{\partial \phi} = -  \hat{\boldsymbol{r}} \qquad \frac{\partial \hat{\boldsymbol{r}}}{\partial r} = \boldsymbol{0} \qquad \frac{\partial \hat{\boldsymbol{\phi}}}{\partial r} =  \boldsymbol{0}.
\end{equation}
However, we have $\hat{\boldsymbol{r}}(r(R,\Phi),\phi(R,\Phi))$ and same for $\hat{\boldsymbol{\phi}}$. Thus, we use the chain rule to get
\begin{equation}\label{eqn:UnitVectorDerivativesNew}
\frac{d \hat{\boldsymbol{r}}}{d R} = \phi_R \hat{\boldsymbol{\phi}} \quad \frac{d \hat{\boldsymbol{r}}}{d \Phi} = \phi_\Phi \hat{\boldsymbol{\phi}} \quad \frac{d \hat{\boldsymbol{\phi}}}{d R} = - \phi_R \hat{\boldsymbol{r}} \quad \frac{d \hat{\boldsymbol{\phi}}}{d \Phi} = -\phi_\Phi \hat{\boldsymbol{r}}.
\end{equation}
Introducing auxiliary functions for the nonlinearities evaluated at (current configuration) strains, as detailed in Eq. \eqref{eqn:DefsOfMathcalK}, we get Eq. \eqref{eqn:ContinuumProblemCurrentUnitVectors}.


\subsection{Pulling back unit vectors to the unloaded configuration}

We first transform from the current configuration $(\hat{\boldsymbol{r}},\hat{\boldsymbol{\phi}})$ to reference configuration $(\hat{\boldsymbol{R}},\hat{\boldsymbol{\Phi}})$. Expressing these unit vectors in Cartesian coordinate system, we get the following coordinates
$$\hat{\boldsymbol{r}} = (\cos{\phi(R,\Phi)}, \sin{\phi(R,\Phi)}) \quad \hat{\boldsymbol{\phi}} = (-\sin{\phi(R,\Phi)}, \cos{\phi(R,\Phi)}) \quad \hat{\boldsymbol{R}} = (\cos{\Phi}, \sin{\Phi}) \quad \hat{\boldsymbol{\Phi}} = (-\sin{\Phi}, \cos{\Phi}) $$
from which using the ansatz
$$\hat{\boldsymbol{r}} = a(R,\Phi) \hat{\boldsymbol{R}} + b(R,\Phi) \hat{\boldsymbol{\Phi}}$$
for unknown functions $a(R,\Phi)$ and $b(R,\Phi)$ and observing that $\hat{\boldsymbol{\phi}}$ can be obtained by rotating $\hat{\boldsymbol{r}}$ by $\pi/2$, we obtain
\begin{equation}\label{eqn:TransformUnitPolarVectorsLatest}
\hat{\boldsymbol{r}} = \cos{(\phi(R,\Phi)-\Phi)} \hat{\boldsymbol{R}} + \sin{(\phi(R,\Phi)-\Phi)} \hat{\boldsymbol{\Phi}} \quad \hat{\boldsymbol{\phi}} = - \sin{(\phi(R,\Phi)-\Phi)} \hat{\boldsymbol{R}} + \cos{(\phi(R,\Phi)-\Phi)} \hat{\boldsymbol{\Phi}}.
\end{equation}
Substituting these relations into governing equations, we regroup (factoring out $\hat{\boldsymbol{R}}$ and $\hat{\boldsymbol{\Phi}}$) and simplifying get 
\begin{equation}\label{eqn:AlmostForceBalanceContinuum}
\begin{aligned}
& \boldsymbol{0} = \Bigg\{ \bigg(  \left(  \mathcal{G}^{(2)} + \mathcal{G}_{+}^{(3)} + \mathcal{G}_{-}^{(3)} \right) \left( r \cos{(\phi-\Phi)} \right)_R  + \kappa \left( \mathcal{G}_{+}^{(3)} - \mathcal{G}_{-}^{(3)} \right) \left( r \cos{(\phi-\Phi)} \right)_\Phi - \kappa \left( \mathcal{G}_{+}^{(3)} - \mathcal{G}_{-}^{(3)} \right)  r \sin{(\phi-\Phi)} \bigg)_R + \\
& \bigg( \kappa \left( \mathcal{G}_{+}^{(3)} - \mathcal{G}_{-}^{(3)} \right) \left( r \cos{(\phi-\Phi)} \right)_R + \kappa^2 \left( \mathcal{G}^{(1)}+ \mathcal{G}_{+}^{(3)} + \mathcal{G}_{-}^{(3)} \right) \left( r \cos{(\phi-\Phi)} \right)_\Phi - \kappa^2 \left( \mathcal{G}^{(1)}+ \mathcal{G}_{+}^{(3)} + \mathcal{G}_{-}^{(3)} \right)  r \sin{(\phi-\Phi)} \bigg)_{\Phi} - \\
& \bigg( \kappa \left( \mathcal{G}_{+}^{(3)} - \mathcal{G}_{-}^{(3)} \right) \left( r \sin{(\phi-\Phi)} \right)_R + \kappa^2 \left( \mathcal{G}^{(1)}+ \mathcal{G}_{+}^{(3)} + \mathcal{G}_{-}^{(3)} \right) \left( r \sin{(\phi-\Phi)} \right)_\Phi + \kappa^2
\left(\mathcal{G}^{(1)}+  \mathcal{G}_{+}^{(3)} + \mathcal{G}_{-}^{(3)} \right)  r \cos{(\phi-\Phi)} \bigg)  \Bigg\} \hat{\boldsymbol{R}} + \\
& \Bigg\{ \bigg( \left(  \mathcal{G}^{(2)} + \mathcal{G}_{+}^{(3)} + \mathcal{G}_{-}^{(3)} \right) \left( r \sin{(\phi-\Phi)} \right)_R + \kappa \left( \mathcal{G}_{+}^{(3)} - \mathcal{G}_{-}^{(3)} \right) \left( r \sin{(\phi-\Phi)} \right)_\Phi + \kappa \left( \mathcal{G}_{+}^{(3)} - \mathcal{G}_{-}^{(3)} \right)  r \cos{(\phi-\Phi)}  \bigg)_R + \\
& \bigg( \kappa \left( \mathcal{G}_{+}^{(3)} - \mathcal{G}_{-}^{(3)} \right) \left( r \sin{(\phi-\Phi)} \right)_R + \kappa^2 \left( \mathcal{G}^{(1)}+ \mathcal{G}_{+}^{(3)} + \mathcal{G}_{-}^{(3)} \right) \left( r \sin{(\phi-\Phi)} \right)_\Phi + \kappa^2
\left(\mathcal{G}^{(1)}+  \mathcal{G}_{+}^{(3)} + \mathcal{G}_{-}^{(3)} \right)  r \cos{(\phi-\Phi)} \bigg)_{\Phi} + \\
& \kappa \left( \mathcal{G}_{+}^{(3)} - \mathcal{G}_{-}^{(3)} \right) \left( r \cos{(\phi-\Phi)} \right)_R + \kappa^2 \left( \mathcal{G}^{(1)}+ \mathcal{G}_{+}^{(3)} + \mathcal{G}_{-}^{(3)} \right) \left( r \cos{(\phi-\Phi)} \right)_\Phi - \kappa^2 \left( \mathcal{G}^{(1)}+ \mathcal{G}_{+}^{(3)} + \mathcal{G}_{-}^{(3)} \right)  r \sin{(\phi-\Phi)} \Bigg\} \hat{\boldsymbol{\Phi}}.
\end{aligned}
\end{equation}
This is satisfied provided both factors multiplying $\hat{\boldsymbol{R}}$ and $\hat{\boldsymbol{\Phi}}$ are $0$, which gives us two equations. However, recall that in the process of deriving these two equations, we divided by $\varepsilon$ twice - first right after Eq. \eqref{eqn:DefOfMu} and then also recall that the first nontrivial balance was found at $O(\varepsilon)$.
Multiplying \eqref{eqn:AlmostForceBalanceContinuum} by $\ve^2$ and redimensionalizing we therefore obtain statements of force balance. 

\subsection{Deducing the nominal stress tensor and strain energy density}

Next we would like to interpret this as $\text{Div} (\boldsymbol{S})=0$ for some two-dimensional nominal stress tensor $\boldsymbol{S}$, which in polar coordinates can be written as
\begin{equation}\label{eqn:DivSInPolarCoord}
\frac{1}{R} \left\{ \left(R S_{RR} \right)_R + S_{\Phi R, \Phi} - S_{\Phi \Phi} \right\} = 0 \qquad \frac{1}{R} \left\{ \left( R S_{R \Phi} \right)_R + S_{\Phi \Phi, \Phi} + S_{\Phi R} \right\}=0.
\end{equation}
To achieve this, we need to divide these forces by the (dimensionless) area of the unit cell, i.e. $\varepsilon^2 \kappa R$, as well as by the (dimensionless) length in the out-of-plane direction $D$, and comparing with \eqref{eqn:DivSInPolarCoord},  we can read off the components of the nominal stress tensor $S$ as
$$S_{RR} = \frac{1}{\kappa R D} \left( \left(  \mathcal{G}^{(2)} + \mathcal{G}_{+}^{(3)} + \mathcal{G}_{-}^{(3)} \right) \left( r \cos{(\phi-\Phi)} \right)_R + \kappa \left( \mathcal{G}_{+}^{(3)} - \mathcal{G}_{-}^{(3)} \right) \left( r \cos{(\phi-\Phi)} \right)_\Phi - \kappa \left( \mathcal{G}_{+}^{(3)} - \mathcal{G}_{-}^{(3)} \right)  r \sin{(\phi-\Phi)} \right) $$
$$S_{R \Phi} = \frac{1}{\kappa R D} \left( \left(  \mathcal{G}^{(2)} + \mathcal{G}_{+}^{(3)} + \mathcal{G}_{-}^{(3)} \right) \left( r \sin{(\phi-\Phi)} \right)_R + \kappa \left( \mathcal{G}_{+}^{(3)} - \mathcal{G}_{-}^{(3)} \right) \left( r \sin{(\phi-\Phi)} \right)_\Phi + \kappa \left( \mathcal{G}_{+}^{(3)} - \mathcal{G}_{-}^{(3)} \right)  r \cos{(\phi-\Phi)} \right) $$
$$S_{\Phi R} = \frac{1}{D} \left( \left( \mathcal{G}_{+}^{(3)} - \mathcal{G}_{-}^{(3)} \right) \left( r \cos{(\phi-\Phi)} \right)_R + \kappa \left( \mathcal{G}^{(1)}+ \mathcal{G}_{+}^{(3)} + \mathcal{G}_{-}^{(3)} \right) \left( r \cos{(\phi-\Phi)} \right)_\Phi - \kappa \left( \mathcal{G}^{(1)}+ \mathcal{G}_{+}^{(3)} + \mathcal{G}_{-}^{(3)} \right)  r \sin{(\phi-\Phi)} \right) $$
$$S_{\Phi \Phi} = \frac{1}{D}  \left( \left( \mathcal{G}_{+}^{(3)} - \mathcal{G}_{-}^{(3)} \right) \left( r \sin{(\phi-\Phi)} \right)_R + \kappa \left( \mathcal{G}^{(1)}+ \mathcal{G}_{+}^{(3)} + \mathcal{G}_{-}^{(3)} \right) \left( r \sin{(\phi-\Phi)} \right)_\Phi + \kappa
\left(\mathcal{G}^{(1)}+  \mathcal{G}_{+}^{(3)} + \mathcal{G}_{-}^{(3)} \right)  r \cos{(\phi-\Phi)}  \right) $$

\subsubsection{Stress tensor as function of displacement field}

We can express the displacement field in both Cartesian and polar coordinates as
$$ (x(X,Y)-X) \hat{\boldsymbol{e}}_{X} + (y(X,Y)-Y) \hat{\boldsymbol{e}}_{Y}  = u(X,Y) = u(R, \Phi) = u^{R} \hat{\boldsymbol{e}}_{R} + u^{\Phi} \hat{\boldsymbol{e}}_{\Phi}. $$
Using the rule for transforming between the coordinate systems we get
\begin{equation}
 \begin{bmatrix}
 u^{R} \\
 u^{\Phi}
\end{bmatrix}{} =
\left[\begin{array}{cc}
  \cos{\Phi} & \sin{\Phi} \\
  - \sin{\Phi} & \cos{\Phi} \\ 
\end{array}\right]
\begin{bmatrix}
  x-X  \\
  y-Y
\end{bmatrix},
\end{equation}
from which it follows
$$ u^R = r \cos{(\phi - \Phi)} - R \qquad u^{\Phi} = r \sin{(\phi-\Phi)} $$
Using simple manipulation we can then write
$$S_{RR} = \frac{1}{\kappa R D } \left( \left( \mathcal{G}^{(2)} + \mathcal{G}_{+}^{(3)} + \mathcal{G}_{-}^{(3)} \right) \left( 1 + u_{R}^{R} \right)  + \kappa \left( \mathcal{G}_{+}^{(3)} - \mathcal{G}_{-}^{(3)} \right) \left( u_{\Phi}^{R} - u^{\Phi}\right) \right) $$
$$S_{R \Phi} = \frac{1}{\kappa R D} \left( \left( \mathcal{G}^{(2)} + \mathcal{G}_{+}^{(3)} + \mathcal{G}_{-}^{(3)} \right) \left( u_{R}^{\Phi} \right)  + \kappa \left( \mathcal{G}_{+}^{(3)} - \mathcal{G}_{-}^{(3)} \right) \left( u_{\Phi}^{\Phi} + u^{R} + R\right) \right)  $$
$$S_{\Phi R} = \frac{1}{D}  \left( \left( \mathcal{G}_{+}^{(3)} - \mathcal{G}_{-}^{(3)} \right) \left( 1 + u_{R}^{R} \right)  + \kappa \left( 
 \mathcal{G}^{(1)}+ \mathcal{G}_{+}^{(3)} + \mathcal{G}_{-}^{(3)} \right) \left( u_{\Phi}^{R} - u^{\Phi}\right)  \right) $$
$$S_{\Phi \Phi} = \frac{1}{D} \left( \left( \mathcal{G}_{+}^{(3)} - \mathcal{G}_{-}^{(3)} \right) \left( u_{R}^{\Phi} \right)  + \kappa \left( 
 \mathcal{G}^{(1)}+ \mathcal{G}_{+}^{(3)} + \mathcal{G}_{-}^{(3)} \right) \left( u_{\Phi}^{\Phi} + u^R +R  \right) \right) $$
where
$$ \mathcal{G}^{(1)}= g^{(1)} \left( \sqrt{\left( \frac{u_{\Phi}^R - u^{\Phi}}{R} \right)^2 + \left( \frac{u_{\Phi}^{\Phi} + u^R +R }{R} \right)^2} \right), \qquad \mathcal{G}^{(2)} = g^{(2)} \left(\sqrt{\left( 1+ u_R^R \right)^2 + \left(u_R^{\Phi} \right)^2} \right)$$
and
$$ \mathcal{G}_{\pm}^{(3)} = g^{(3)} \left( \sqrt{ \frac{ \displaystyle \left( 1 + u_{R}^R \pm \kappa R \left( \frac{u_{\Phi}^R - u^{\Phi}}{R} \right) \right)^2 + \left( u_{R}^{\Phi} \pm \kappa R \left( \frac{u_{\Phi}^{\Phi} + u^R + R}{R} \right) \right)^2}{\displaystyle 1 + \kappa^2 R^2} } \right). $$
We can next express the (transpose of the) deformation gradient tensor in polar coordinates as
\begin{equation}\label{eqn:DefGradTensorTransportPolar}
\boldsymbol{F}^T = \boldsymbol{I} + \left( \nabla u \right)^{T} = \left[\begin{array}{cc}
  1+u_{R}^R & u_{R}^{\Phi} \\
  \displaystyle \frac{u_{\Phi}^R - u^{\Phi}}{R} & \displaystyle \frac{u_{\Phi}^{\Phi} + u^R + R}{R} \\ 
\end{array}\right],
\end{equation}
and introducing the right Cauchy-Green tensor $\boldsymbol{C} = \boldsymbol{F}^T \boldsymbol{F}$, we find
$$ \mathcal{G}^{(1)}= g^{(1)} \left( \sqrt{\hat{\boldsymbol{e}}_\Phi \cdot \left(\boldsymbol{C} \hat{\boldsymbol{e}}_\Phi \right)} \right), \qquad \mathcal{G}^{(2)} = g^{(2)} \left(\sqrt{\hat{\boldsymbol{e}}_R \cdot \left(\boldsymbol{C} \hat{\boldsymbol{e}}_R \right)} \right) \qquad \text{and} \qquad  \mathcal{G}_{\pm}^{(3)} = g^{(3)} \left( \sqrt{\hat{\boldsymbol{e}}_{\pm} \cdot \left( \boldsymbol{C} \hat{\boldsymbol{e}}_{\pm} \right) } \right), $$
where $\hat{\boldsymbol{e}}_{R}$, $\hat{\boldsymbol{e}}_{\Phi}$ and $\hat{\boldsymbol{e}}_{\pm}$ are the unit normal vectors pointing along the directions of the radial cross-links, lamellar segments and diagonal cross-links, respectively. In polar coordinates, these vectors take form
$$ \hat{\boldsymbol{e}}_{R} = (1,0), \qquad   \hat{\boldsymbol{e}}_{\Phi} = (0,1), \qquad \text{and} \qquad \hat{\boldsymbol{e}}_{\pm} = \left( \frac{1}{\sqrt{1+ \kappa^2 R^2}}, \pm \frac{\kappa R}{\sqrt{1+ \kappa^2 R^2}} \right).$$

\subsubsection{Redimensionalization}

In order to redimensionalize the stress tensor, we first recall that $\kappa$ was defined as the quotient of the dimensionless steps in the angular and radial directions, i.e.
$$\kappa = \frac{\varepsilon_{\Phi}}{\varepsilon}$$
and its redimensionalized counterpart thus reads
$$\tilde{\kappa} = \frac{\varepsilon_{\Phi}}{\varepsilon \tilde{T}} = \frac{\kappa}{\tilde{T}}.$$
The redimensionalized nominal stress tensor $\tilde{\boldsymbol{S}}$ in polar coordinates reads
\begin{equation}\label{eqn:UpscaledDimensionalStressTensor}
\begin{aligned}
& \left[\begin{array}{cc}
    \displaystyle \frac{ \left( \tilde{\mathcal{G}}^{(2)} + \tilde{\mathcal{G}}_{+}^{(3)} + \tilde{\mathcal{G}}_{-}^{(3)} \right) \left(1 + \tilde{u}_{\tilde{R}}^{\tilde{R}} \right) + \tilde{\kappa} \left( \tilde{\mathcal{G}}_{+}^{(3)} - \tilde{\mathcal{G}}_{-}^{(3)} \right) \left( \tilde{u}_{\Phi}^{\tilde{R}} - \tilde{u}^{\Phi} \right)  }{\tilde{\kappa} \tilde{R} \tilde{D}} & \displaystyle \frac{ \left( \tilde{\mathcal{G}}^{(2)} + \tilde{\mathcal{G}}_{+}^{(3)} + \tilde{\mathcal{G}}_{-}^{(3)} \right) \left( \tilde{u}_{\tilde{R}}^{\Phi} \right) + \tilde{\kappa} \left( \tilde{\mathcal{G}}_{+}^{(3)} - \tilde{\mathcal{G}}_{-}^{(3)} \right) \left( \tilde{u}_{\Phi}^{\Phi} + \tilde{u}^{\tilde{R}} + \tilde{R} \right)  }{ \tilde{\kappa} \tilde{R} \tilde{D}} \\[12pt]
  \displaystyle  \frac{\left( \tilde{\mathcal{G}}_{+}^{(3)} - \tilde{\mathcal{G}}_{-}^{(3)} \right) \left(1 + \tilde{u}_{\tilde{R}}^{\tilde{R}} \right) + \tilde{\kappa} \left( \tilde{\mathcal{G}}^{(1)} +  \tilde{\mathcal{G}}_{+}^{(3)} + \tilde{\mathcal{G}}_{-}^{(3)} \right) \left( \tilde{u}_{\Phi}^{\tilde{R}} - \tilde{u}^{\Phi} \right)}{\tilde{D}}  &  \displaystyle  \frac{\left( \tilde{\mathcal{G}}_{+}^{(3)} - \tilde{\mathcal{G}}_{-}^{(3)} \right) \left( \tilde{u}_{\tilde{R}}^{\Phi} \right) + \tilde{\kappa} \left( \tilde{\mathcal{G}}^{(1)} + \tilde{\mathcal{G}}_{+}^{(3)} + \tilde{\mathcal{G}}_{-}^{(3)} \right) \left( \tilde{u}_{\Phi}^{\Phi} + \tilde{u}^{\tilde{R}} + \tilde{R} \right)}{\tilde{D}} \\ 
\end{array}\right]\\
& = \frac{1}{\tilde{D}} \left[\begin{array}{cc} 
  \left( \tilde{\mathcal{G}}^{(2)} + \tilde{\mathcal{G}}_{+}^{(3)} + \tilde{\mathcal{G}}_{-}^{(3)} \right)/\left(\tilde{R} \tilde{\kappa} \right) &  \tilde{\mathcal{G}}_{+}^{(3)} - \tilde{\mathcal{G}}_{-}^{(3)} \\
 \tilde{\mathcal{G}}_{+}^{(3)} - \tilde{\mathcal{G}}_{-}^{(3)} & \tilde{\kappa} \tilde{R} \left( \tilde{\mathcal{G}}^{(1)} + \tilde{\mathcal{G}}_{+}^{(3)} + \tilde{\mathcal{G}}_{-}^{(3)} \right) \\ 
\end{array}\right] \boldsymbol{F}^T = \tilde{\boldsymbol{T}} \boldsymbol{F}^T,
\end{aligned}
\end{equation}
which introduces the second Piola-Kirchhoff tensor $\tilde{\boldsymbol{T}}$ and confirms its symmetry. Assuming all structural elements behave as linear springs we obtain Eq. \eqref{eqn:SecondPiolaKirchhoff}. It is similarly straight-forward to deduce the Cauchy stress tensor and show that it is symmetric. For completeness, we summarize the governing equations of the dimensional continuum model under the linear-springs assumption as follows. One must solve
\begin{subequations}\label{eqn:FullDimensionalContinuumModel}
\begin{equation}\label{eqn:FullDimensionalContinuumModel_Interior}
\tilde{\nabla} \cdot \tilde{\boldsymbol{S}} = \boldsymbol{0}
\end{equation}
in the corneal interior ($\tilde{R}_P < \tilde{R} < \tilde{R}_A$ and $-\Phi^* <  \Phi < \Phi^*$), subject to
\begin{equation}\label{eqn:FullDimensionalContinuumModel_Limbus}
\tilde{\boldsymbol{x}} = \tilde{\boldsymbol{X}} 
\end{equation}
at the limbus ($\Phi = \pm \Phi^*$), 
\begin{equation}\label{eqn:FullDimensionalContinuumModel_Posterior}
\tilde{\boldsymbol{S}}^T \boldsymbol{N} = - \tilde{p} \text{ det}(\boldsymbol{F}) \boldsymbol{F}^{-T} \boldsymbol{N}
\end{equation}
at the posterior ($\tilde{R} = \tilde{R}_P$; see Section~\ref{eqn:SuppSubSec_ContinuumBCs} for details) and
\begin{equation}\label{eqn:FullDimensionalContinuumModel_Anterior}
\tilde{\boldsymbol{S}}^T \boldsymbol{N} = \boldsymbol{0}
\end{equation}
at the anterior ($\tilde{R} = \tilde{R}_A$) boundary, where $\boldsymbol{N}$ is the unit vector normal to the boundary and pointing outside the cornea, $\tilde{p}$ denotes the dimensional intraocular pressure and
\begin{equation}\label{eqn:ContinuumDimensional_RelateSAndT}
\tilde{\boldsymbol{S}} = \tilde{\boldsymbol{T}} \boldsymbol{F}^T
\end{equation}
\end{subequations}
relates the nominal stress tensor $\tilde{\boldsymbol{S}}$ to the second Piola-Kirchhoff stress tensor $\tilde{\boldsymbol{T}}$ defined in \eqref{eqn:SecondPiolaKirchhoff}. Defining
$$ \tilde{e}^{(1)}(\lambda) = \frac{1}{\tilde{D}} \int\limits_{1}^{\lambda} y \tilde{g}^{(1)} (y) dy, \qquad \tilde{e}^{(2)}(\lambda) = \frac{1}{\tilde{D}}\int\limits_{1}^{\lambda} y \tilde{g}^{(2)} (y) dy \qquad \text{and} \qquad \tilde{e}^{(3)}(\lambda) = \frac{1}{\tilde{D}} \int\limits_{1}^{\lambda} y \tilde{g}^{(3)} (y) dy,$$
we deduce the dimensional strain-energy density using the relationship $ \tilde{S}_{ij} = \partial \tilde{W}/\partial F_{ji} $ and get
\begin{equation}\label{eqn:FinalStrainEnergy}
\tilde{W} = \frac{\tilde{\kappa}^2 \tilde{R}^2}{\tilde{\kappa} \tilde{R}} \tilde{e}^{(1)} \left( \sqrt{ \hat{\boldsymbol{e}}_\Phi \cdot \left(\boldsymbol{C} \hat{\boldsymbol{e}}_\Phi \right) }  \right) +  \frac{1}{ \tilde{\kappa} \tilde{R}} \tilde{e}^{(2)} \left(\sqrt{ \hat{\boldsymbol{e}}_R \cdot \left(\boldsymbol{C} \hat{\boldsymbol{e}}_R \right) } \right) + \frac{1+ \tilde{\kappa}^2 \tilde{R}^2}{ \tilde{\kappa} \tilde{R}} \tilde{e}^{(3)} \left( \sqrt{ \hat{\boldsymbol{e}}_+ \cdot \left(\boldsymbol{C} \hat{\boldsymbol{e}}_+ \right) }  \right) +  \frac{1+ \tilde{\kappa}^2 \tilde{R}^2}{ \tilde{\kappa} \tilde{R}} \tilde{e}^{(3)} \left(\sqrt{ \hat{\boldsymbol{e}}_- \cdot \left(\boldsymbol{C} \hat{\boldsymbol{e}}_- \right) } \right),
\end{equation}
where the numerators and denominators can be interpreted as the squares of side (or diagonal) lengths and area of the unit cell in the continuum limit, respectively (see Fig. \ref{fig:ReferenceGeometryDiscretized}). Assuming all elements behave as linear springs we get
$$  \tilde{e}^{(m)} (\lambda) = \frac{\tilde{K}^{(m)}}{2 \tilde{D}} \left( \lambda - 1 \right)^2,$$ 
which yields the strain-energy density \eqref{eqn:StrainEnergyLinearSprings}. In polar coordinates, we can express \eqref{eqn:StrainEnergyLinearSprings} also as
\begin{equation}\label{eqn:StrainEnergyLinearSprings2}
\begin{aligned}
\tilde{W} = 
& \frac{1}{2 \kappa R \tilde{D}}  \left\{ \left( \kappa R \right)^2 \tilde{K}^{(1)} \left( \sqrt{C_{\Phi \Phi}} - 1 \right)^2 + \tilde{K}^{(2)}  \left( \sqrt{C_{R R}} - 1 \right)^2 + \right. \\
& \left. 2 \tilde{K}^{(3)} \left[
1 + C_{RR} + (1 + C_{\Phi\Phi}) (\kappa R)^2 -  \sqrt{1 + (\kappa R)^2} \right. \right. \\
&
\left. \left. \left(
\sqrt{C_{RR} + 2C_{\Phi R} \kappa R + C_{\Phi\Phi} (\kappa R)^2 } +
\sqrt{C_{RR} - 2C_{\Phi R} \kappa R + C_{\Phi\Phi} (\kappa R)^2 } 
\right) \right] \right\}.
\end{aligned}
\end{equation}

\subsection{Strain energy under small deformations: relation to linear elasticity}
\label{sec:Appendix_StrainEnergySmallDef}
Writing the loaded-configuration variables as a small perturbation of the unloaded-configuration variables:
$$ \boldsymbol{x} = \boldsymbol{X} + \boldsymbol{u}(\boldsymbol{X}) = \boldsymbol{X} + \delta \boldsymbol{x}'(\boldsymbol{X}) + \delta^2 \boldsymbol{x}''(\boldsymbol{X}) + O(\delta^3) , \hspace{0.5cm} \text{where} \hspace{0.2cm} \delta \ll 1, \, \boldsymbol{x}' = O(1) \text{ and } \boldsymbol{x}'' = O(1) ,$$
we find that
\begin{equation}\label{eqn:IncludesDefnOfE}
\boldsymbol{C} = \boldsymbol{I} + 2 \boldsymbol{E} = \boldsymbol{I} + \delta \boldsymbol{A} + \delta^2 \boldsymbol{B} + O(\delta^3),
\end{equation}
where $\boldsymbol{I}$ is an identity tensor, 
$$\boldsymbol{E} = \frac{1}{2} \left[ \nabla_X \boldsymbol{u} + (\nabla_X \boldsymbol{u})^T + (\nabla_X \boldsymbol{u})^T \cdot \nabla_X \boldsymbol{u} \right]$$ 
is the Lagrangian finite strain tensor, and $O(1)$ tensors $\boldsymbol{A}$ and $\boldsymbol{B}$ are given by
$$ \boldsymbol{A} =  \nabla_X \boldsymbol{x}' + \left( \nabla_X \boldsymbol{x}' \right)^T \qquad \boldsymbol{B} = \nabla_X \boldsymbol{x}'' + \left( \nabla_X \boldsymbol{x}'' \right)^T + \left( \nabla_X \boldsymbol{x}' \right)^T  \cdot \nabla_X \boldsymbol{x}'.$$
Therefore, we can write \eqref{eqn:StrainEnergyLinearSprings2} as
\begin{equation}\label{eqn:StrainEnergyLinearSprings3}
\begin{aligned}
& \tilde{W} = 
 \frac{1}{2 \kappa R \tilde{D}}  \left\{ \left( \kappa R \right)^2 \tilde{K}^{(1)} \left( \sqrt{1+ \delta A_{\Phi \Phi} + O(\delta^2)} - 1 \right)^2 + \tilde{K}^{(2)}  \left( \sqrt{1 + \delta A_{RR}+ O(\delta^2)} - 1 \right)^2 + \right. \\
& \left. 2 \tilde{K}^{(3)} \left[
2 + \delta A_{RR} + \delta^2 B_{RR} + O(\delta^3) + (2 + \delta A_{\Phi\Phi} + \delta^2 B_{\Phi \Phi} + O(\delta^3)) (\kappa R)^2 -  \sqrt{1 + (\kappa R)^2} \right. \right. \\
&
\left. \left. \left(
\sqrt{ 1 + (\kappa R)^2 + \delta \left( A_{RR} + 2 \kappa R A_{R \Phi} +(\kappa R)^2 A_{\Phi \Phi}  \right) + \delta^2 ( B_{RR} + 2 \kappa R B_{R \Phi} +(\kappa R)^2 B_{\Phi \Phi}) + O(\delta^3)}  +  \right. \right. \right. \\
& \left. \left. \left.
\sqrt{ 1 + (\kappa R)^2 + \delta \left( A_{RR} - 2 \kappa R A_{R \Phi} +(\kappa R)^2 A_{\Phi \Phi}  \right) + \delta^2 ( B_{RR} - 2 \kappa R B_{R \Phi} +(\kappa R)^2 B_{\Phi \Phi}) + O(\delta^3)} 
\right) \right] \right\}.
\end{aligned}
\end{equation}


Grouping together different exponents of $\delta$ and Taylor expanding for $\delta \ll 1$ we get
\begin{equation}\label{eqn:StrainEnergyLinearSprings4}
\begin{aligned}
& \tilde{W} = 
 \frac{1}{2 \kappa R \tilde{D}}  \left\{ \left( \kappa R \right)^2 \tilde{K}^{(1)} \left( \frac{1}{4} A_{\Phi \Phi}^2 \delta^2 + O(\delta^3) \right) + \tilde{K}^{(2)}  \left( \frac{1}{4} A_{RR}^2 \delta^2 + O(\delta^3) \right) + 2 \tilde{K}^{(3)} \Bigg[ 2 \left( 1 + (\kappa R)^2 \right) + \delta \left( A_{RR} + (\kappa R)^2 A_{\Phi \Phi} \right) + \right. \\
& \left. \delta^2 \left( B_{RR} + (\kappa R)^2 B_{\Phi \Phi} \right)  -  \sqrt{1 + (\kappa R)^2} \left(
\sqrt{1+ (\kappa R)^2} + \delta \frac{A_{RR} + 2 \kappa R A_{R \Phi} +(\kappa R)^2 A_{\Phi \Phi} }{2 \sqrt{1+ (\kappa R)^2}} + \right.  \right.  \\
&
\left. \left. \delta^2 \frac{4  \left( 1+ (\kappa R)^2 \right) \left( B_{RR} + 2 \kappa R B_{R \Phi} +(\kappa R)^2 B_{\Phi \Phi}) \right) - \left( A_{RR} + 2 \kappa R A_{R \Phi} +(\kappa R)^2 A_{\Phi \Phi}  \right)^2}{8 \left( 1+ (\kappa R)^2 \right)^{3/2} } + O(\delta^3) + \sqrt{1+ (\kappa R)^2} +  \right. \right. \\
&  \left. \left. \delta \frac{A_{RR} - 2 \kappa R A_{R \Phi} +(\kappa R)^2 A_{\Phi \Phi} }{2 \sqrt{1+ (\kappa R)^2}} +
  \delta^2 \frac{4  \left( 1+ (\kappa R)^2 \right) \left( B_{RR} - 2 \kappa R B_{R \Phi} +(\kappa R)^2 B_{\Phi \Phi}) \right) - \left( A_{RR} - 2 \kappa R A_{R \Phi} +(\kappa R)^2 A_{\Phi \Phi}  \right)^2}{8 \left( 1+ (\kappa R)^2 \right)^{3/2} } 
\right) \Bigg] \right\}.
\end{aligned}
\end{equation}


All contributions at $O(1)$ and $O(\delta)$ cancel out and the first nontrivial contribution comes at $O(\delta^2)$. At this order, all contributions from $\boldsymbol{B}$ tensor vanish and we conclude
\begin{equation}\label{eqn:StrainEnergyLinearSpringsSimplified}
\tilde{W} = \frac{\delta^2}{8 \kappa R \tilde{D}} \left\{ (\kappa R)^2 \tilde{K}^{(1)} A_{\Phi \Phi}^2 + \tilde{K}^{(2)} A_{R R}^2 + \frac{2 \tilde{K}^{(3)}}{1+(\kappa R)^2} \left( \left(A_{RR} + (\kappa R)^2 A_{\Phi \Phi} \right)^2 + 4 (\kappa R)^2 A_{R \Phi}^2 \right)    \right\} + O(\delta^3).
\end{equation}
An alternative derivation of the above expression is provided in Sec. \ref{sec:Appendix_AltDerivationStrainEnergySmallDef}. Realizing that $1/2 \delta \boldsymbol{A} = \boldsymbol{\epsilon}$, where $\boldsymbol{\epsilon}$ is an infinitesimal strain tensor, the leading-order strain energy can be rewritten as
\begin{equation}\label{eqn:StrainEnergyLinearSpringsSimplified2}
\tilde{W} = \frac{1}{2 \kappa R \tilde{D}} \left\{ (\kappa R)^2 \tilde{K}^{(1)} \epsilon_{\Phi \Phi}^2 + \tilde{K}^{(2)} \epsilon_{R R}^2 + \frac{2 \tilde{K}^{(3)}}{1+(\kappa R)^2} \left( \left(\epsilon_{RR} + (\kappa R)^2 \epsilon_{\Phi \Phi} \right)^2 + 4 (\kappa R)^2 \epsilon_{R \Phi}^2 \right)    \right\}  + O(\delta^3).
\end{equation}
At the leading order, the strain energy can then be rewritten as
\begin{equation}\label{eqn:StrainEnergyStressStrain}
\tilde{W} = 
\frac{1}{2} 
\left( 
\tilde{\sigma}_{R R}  \epsilon_{R R} + \tilde{\sigma}_{\Phi \Phi} \epsilon_{\Phi \Phi} + 2 \tilde{\sigma}_{R \Phi} \epsilon_{R\Phi} \right),
\end{equation}
where
\begin{equation}\label{eqn:StressAsFtionOfStrainLinear}
\begin{aligned}
\tilde{\sigma}_{RR} & = \frac{1}{\tilde{D}} \left\{ \tilde{K}^{(2)} \epsilon_{RR} + \frac{2 \tilde{K}^{(3)}}{1 + (\kappa R)^2} \left( \epsilon_{RR} + (\kappa R)^2 \epsilon_{\Phi \Phi} \right) \right\} \\
\tilde{\sigma}_{\Phi\Phi} & = \frac{1}{\tilde{D}} \left\{  (\kappa R)^2 \tilde{K}^{(1)} \epsilon_{\Phi \Phi} + \frac{2 \tilde{K}^{(3)}}{1 + (\kappa R)^2} \left( (\kappa R)^2 \epsilon_{RR} + (\kappa R)^4 \epsilon_{\Phi \Phi} \right) \right\}\\
\tilde{\sigma}_{R \Phi} & = \frac{1}{\tilde{D}} \left\{  \frac{4 \tilde{K}^{(3)} (\kappa R)^2}{1 + (\kappa R)^2}  \epsilon_{R\Phi} \right\}. 
\end{aligned}
\end{equation}
Following \cite{Trageser2019Supp} we define the fourth-order elasticity tensor $\tilde{\mathbb{C}}$ (in two dimensions) via
\begin{equation}\label{eqn:DefOfElasticityTensor}
 \left[\begin{array}{c}
  \tilde{\sigma}_{RR} \\[12pt] \tilde{\sigma}_{\Phi \Phi} \\[12pt] \tilde{\sigma}_{R \Phi} \\
 \end{array}\right] 
 =  \left[\begin{array}{ccc}
  \tilde{C}_{RRRR} & \tilde{C}_{RR \Phi \Phi} & \tilde{C}_{RRR \Phi} \\[12pt] \tilde{C}_{RR \Phi \Phi} & \tilde{C}_{\Phi \Phi \Phi \Phi} & \tilde{C}_{\Phi \Phi R \Phi}  \\[12pt] \tilde{C}_{RRR \Phi} & \tilde{C}_{\Phi \Phi R \Phi} & \tilde{C}_{R \Phi R \Phi}  \\
 \end{array}\right] 
  \left[\begin{array}{c}
  \epsilon_{RR} \\[12pt] \epsilon_{\Phi \Phi} \\[12pt] 2 \epsilon_{R \Phi} \\
 \end{array}\right] 
\end{equation}
and conclude
\begin{equation}\label{eqn:FormOfElasticityTensor}
\tilde{\mathbb{C}} = \left[\begin{array}{ccc}
  \tilde{C}_{RRRR} & \tilde{C}_{RR \Phi \Phi} & \tilde{C}_{RRR \Phi} \\[12pt] \tilde{C}_{RR \Phi \Phi} & \tilde{C}_{\Phi \Phi \Phi \Phi} & \tilde{C}_{\Phi \Phi R \Phi}  \\[12pt] \tilde{C}_{RRR \Phi} & \tilde{C}_{\Phi \Phi R \Phi} & \tilde{C}_{R \Phi R \Phi}  \\
 \end{array}\right] = \frac{1}{\tilde{D}}
 \left[\begin{array}{ccc}
  \displaystyle \tilde{K}^{(2)} + \frac{2 \tilde{K}^{(3)}}{1 + (\kappa R)^2} & \displaystyle \frac{2 \tilde{K}^{(3)} (\kappa R)^2}{1+(\kappa R)^2} & 0 \\[12pt] \displaystyle \frac{2 \tilde{K}^{(3)} (\kappa R)^2}{1+(\kappa R)^2} & \displaystyle (\kappa R)^2 \tilde{K}^{(1)} + \frac{2 \tilde{K}^{(3)} (\kappa R)^4}{1 + (\kappa R)^2} & 0  \\[12pt] 0 & 0 & \displaystyle \frac{2 \tilde{K}^{(3)} (\kappa R)^2}{1+(\kappa R)^2}  \\
 \end{array}\right].
 \end{equation}
Note first that \eqref{eqn:StrainEnergyStressStrain} with \eqref{eqn:StressAsFtionOfStrainLinear} is a unique re-writing of (the leading order of) \eqref{eqn:StrainEnergyLinearSpringsSimplified2} such that the resulting elasticity tensor \eqref{eqn:FormOfElasticityTensor} is symmetric. As a check, we further confirm following \cite{Trageser2019Supp} that the resulting elasticity tensor \eqref{eqn:FormOfElasticityTensor} is in the form corresponding to a material belonging to a rectangular symmetry group (with two lines of reflection symmetry) and that the Cauchy relation $\tilde{C}_{RR \Phi \Phi}= \tilde{C}_{R \Phi R \Phi}$ is satisfied - these relations pertain whenever microscale interactions are based on a pairwise potential formulation.


\subsection{Alternative derivation of strain energy assuming small deformations}
\label{sec:Appendix_AltDerivationStrainEnergySmallDef}
As a consistency check, we note that we can also arrive at the same expression for strain energy under small deformation by noting that the second Piola-Kirchhoff stress tensor is work-conjugate tensor to the Lagrangian strain tensor $\boldsymbol{E}$ from \eqref{eqn:IncludesDefnOfE}. We express the Lagrangian strain tensor in terms of $\boldsymbol{A}$ and $\boldsymbol{B}$ as 
$$\boldsymbol{E} = \frac{1}{2} \left(\delta \boldsymbol{A} + \delta^2 \boldsymbol{B} + O(\delta^3) \right).$$
Next, using the linear springs assumption and the smallness of $\delta$, we get
$$\tilde{\mathcal{G}}^{(1)} = \frac{\tilde{K}^{(1)} A_{\Phi \Phi} \delta }{2} + O(\delta^2) \qquad \tilde{\mathcal{G}}^{(2)} = \frac{\tilde{K}^{(2)} A_{R R} \delta }{2} + O(\delta^2) \qquad \tilde{\mathcal{G}}_{\pm}^{(3)} = \frac{\tilde{K}^{(3)} \left( A_{R R}  \pm 2 \tilde{\kappa} \tilde{R} A_{R \Phi} + \left( \tilde{\kappa} \tilde{R} \right)^2 A_{\Phi \Phi} \right) \delta }{2 \left(1 + \left( \tilde{\kappa} \tilde{R} \right)^2 \right)} + O(\delta^2), $$
which when substituted into \eqref{eqn:UpscaledDimensionalStressTensor} gives the second Piola-Kirchhoff stress tensor in polar coordinates
\begin{equation}\label{eqn:PK2_SmallDeform}
\tilde{\boldsymbol{T}} = \frac{\delta}{\tilde{D}} \left[\begin{array}{cc}
 \displaystyle \frac{1}{2 \tilde{\kappa} \tilde{R}} \left( \tilde{K}^{(2)} A_{RR} + \frac{ 2 \tilde{K}^{(3)} \left( A_{RR} + \left( \tilde{\kappa} \tilde{R} \right)^2 A_{\Phi \Phi} \right)}{1 + \left(\tilde{\kappa} \tilde{R}  \right)^2} \right)  & \displaystyle \frac{2 \tilde{K}^{(3)} \tilde{\kappa} \tilde{R} A_{R \Phi} }{1 + \left( \tilde{\kappa} \tilde{R} \right)^2} \\
 \displaystyle \frac{2 \tilde{K}^{(3)} \tilde{\kappa} \tilde{R} A_{R \Phi}}{1 + \left( \tilde{\kappa} \tilde{R} \right)^2} & \displaystyle \frac{\tilde{\kappa} \tilde{R} }{2} \left(\tilde{K}^{(1)} A_{\Phi \Phi} + \frac{ 2 \tilde{K}^{(3)} \left( A_{RR} + \left( \tilde{\kappa} \tilde{R} \right)^2 A_{\Phi \Phi} \right)}{1 + \left(\tilde{\kappa} \tilde{R}  \right)^2} \right) \\ 
\end{array}\right] + O(\delta^2)
\end{equation}
We can then deduce the strain energy via a double contraction
$$ \tilde{W} = \frac{1}{2} \tilde{\boldsymbol{T}} : \boldsymbol{E}$$
which gives \eqref{eqn:StrainEnergyLinearSpringsSimplified}.

\subsection{Continuum boundary conditions}
\label{eqn:SuppSubSec_ContinuumBCs}
At the limbus ($\Phi = \pm \Phi^{*}$) the boundary is pinned and thus we must impose
$$ \tilde{u}^{\tilde{R}} = 0  \qquad \tilde{u}^{\Phi} = 0.$$

\paragraph{Posterior boundary}

Following the same steps as before, we consider a discrete force balance at a node belonging to the posterior boundary (which includes the IOP terms), define the dimensionless intra-occular pressure and out-of-plane thickness as
$$ p = \frac{\tilde{T} \tilde{p}}{\tilde{K}^{(1)}} \qquad D = \frac{\tilde{D}}{\tilde{T}}$$
and get at the leading order in Cartesian coordinates
$$ \varepsilon \left(\mathcal{G}^{(2)} \left(x_R, y_R \right) + \mathcal{G}_{-}^{(3)} \left( x_R - \kappa x_{\Phi}, y_R - \kappa y_{\Phi} \right) + \mathcal{G}_{+}^{(3)} \left( x_R + \kappa x_{\Phi}, y_R + \kappa y_{\Phi} \right) +  \kappa D p \left( y_{\Phi}, -x_{\Phi} \right) \right) =0. $$
Transforming to polar coordinates, we get
$$ \ve \left( \mathcal{G}^{(2)} \left( r_R \hat{\boldsymbol{r}} + r \phi_R \hat{\boldsymbol{\phi}} \right) + \mathcal{G}_{-}^{(3)} \left( \left(r_R - \kappa r_\Phi \right) \hat{\boldsymbol{r}} + \left(r \phi_R - \kappa r \phi_{\Phi} \right) \hat{\boldsymbol{\phi}} \right) + \mathcal{G}_{+}^{(3)} \left( \left(r_R + \kappa r_\Phi \right) \hat{\boldsymbol{r}} + \left(r \phi_R + \kappa r \phi_{\Phi} \right) \hat{\boldsymbol{\phi}} \right) +  \right.$$
$$ \left. \kappa D p \left( r \phi_\Phi \hat{\boldsymbol{r}} - r_\Phi \hat{\boldsymbol{\phi}} \right) \right) = 0, $$
and upon transforming to reference unit vectors, 
we conclude two equations which when expressed in terms of the components of the displacement field read
$$ 0 = \ve \left( \mathcal{G}^{(2)} \left( 1 + u_R^R \right) + \mathcal{G}_{+}^{(3)} \left( 1+u_R^R + \kappa u_{\Phi}^R - \kappa u^{\Phi} \right) +  \mathcal{G}_{-}^{(3)} \left( 1+u_R^R - \kappa u_{\Phi}^R + \kappa u^{\Phi} \right) +  \kappa D p \left( u_{\Phi}^{\Phi} + u^R + R \right)   \right) $$
$$ 0 = \ve \left( \mathcal{G}^{(2)} u_R^{\Phi} + \mathcal{G}_{+}^{(3)} \left( u_R^{\Phi} + \kappa u_{\Phi}^{\Phi} + \kappa \left( u^{R} + R \right) \right) +  \mathcal{G}_{-}^{(3)} \left( u_R^{\Phi} - \kappa u_{\Phi}^{\Phi} - \kappa \left( u^{R} + R \right) \right) +  \kappa D p \left( - u_{\Phi}^{R} + u^{\Phi} \right)   \right).$$
We next divide by the area of the corresponding section of the posterior boundary, which equals its length $\ve \kappa R$ (at $R= R_P$) times the out-of-plane depth $D$, redimensionalize and rewrite as
$$  - \frac{\tilde{\mathcal{G}}^{(2)}}{\tilde{\kappa} \tilde{R} \tilde{D} } \left( 1 + \tilde{u}_{\tilde{R}}^{\tilde{R}} \right) -  \frac{\tilde{\mathcal{G}}_{+}^{(3)}}{\tilde{\kappa} \tilde{R} \tilde{D}} \left( 1+\tilde{u}_{\tilde{R}}^{\tilde{R}} + \tilde{\kappa} \tilde{u}_{\Phi}^{\tilde{R}} - \tilde{\kappa} \tilde{u}^{\Phi} \right) -  \frac{\tilde{\mathcal{G}}_{-}^{(3)}}{\tilde{\kappa} \tilde{R} \tilde{D}} \left( 1+\tilde{u}_{\tilde{R}}^{\tilde{R}} - \tilde{\kappa} \tilde{u}_{\Phi}^{\tilde{R}} + \tilde{\kappa} \tilde{u}^{\Phi} \right) =  \frac{ \tilde{p}}{\tilde{R}} \left( \tilde{u}_{\Phi}^{\Phi} + \tilde{u}^{\tilde{R}} + \tilde{R} \right)   $$
$$ - \frac{\tilde{\mathcal{G}}^{(2)}}{\tilde{\kappa} \tilde{R} \tilde{D}} \tilde{u}_{\tilde{R}}^{\Phi} - \frac{\tilde{\mathcal{G}}_{+}^{(3)}}{\tilde{\kappa} \tilde{R} \tilde{D}} \left( \tilde{u}_{\tilde{R}}^{\Phi} + \tilde{\kappa} \tilde{u}_{\Phi}^{\Phi} + \tilde{\kappa} \left( \tilde{u}^{\tilde{R}} + \tilde{R} \right) \right) - \frac{\tilde{\mathcal{G}}_{-}^{(3)}}{\tilde{\kappa} \tilde{R} \tilde{D}} \left( \tilde{u}_{\tilde{R}}^{\Phi} - \tilde{\kappa} \tilde{u}_{\Phi}^{\Phi} - \tilde{\kappa} \left( \tilde{u}^{\tilde{R}} + \tilde{R} \right) \right) = \frac{ \tilde{p}}{\tilde{R}} \left( - \tilde{u}_{\Phi}^{\tilde{R}} + \tilde{u}^{\Phi} \right).$$
This can be rewritten using the nominal stress tensor $\tilde{\boldsymbol{S}}$ from \eqref{eqn:UpscaledDimensionalStressTensor} and $\tilde{\boldsymbol{F}}^{T}$ from \eqref{eqn:DefGradTensorTransportPolar} as
$$ \tilde{\boldsymbol{S}}^T \boldsymbol{N}  = - \tilde{p}  \text{ adj}(\boldsymbol{F}^T)  \boldsymbol{N}$$
where $\boldsymbol{N} = - \boldsymbol{e}_{R}$ is a unit vector normal to the posterior boundary and pointing outside the cornea and adj() denotes the adjoint operator. Using the formula for inverse matrix, we have
$$ \text{adj}(\boldsymbol{F}^T) = \text{ det}(\boldsymbol{F}^T) \boldsymbol{F}^{-T} = \text{ det}(\boldsymbol{F}) \boldsymbol{F}^{-T} ,$$
where we also used the fact that the determinants of a matrix and its transpose are equal. Finally, we use a connection from nonlinear elasticity between the nominal and the Cauchy stress tensors of the form
$$ \tilde{\boldsymbol{S}}^T = \tilde{\boldsymbol{\sigma}} \text{det}(\boldsymbol{F}) \boldsymbol{F}^{-T} $$
and conclude that the deduced boundary condition is consistent with the boundary condition in the loaded configuration of the form
$$ \tilde{\boldsymbol{\sigma}} = - \tilde{p} \boldsymbol{I}.$$

\end{adjustwidth}

\bibliography{ReferencesSupplementary} 
\bibliographystyle{plain} 

\makeatletter\@input{xx.tex}\makeatother